\documentclass[11pt,twoside]{article} 
\usepackage{cspm-asp2006}
\usepackage{epsfig,graphicx,natbib,url}  
\usepackage{lscape} 
\begin{document}

\markboth{Berlicki}{Chromospheric Line Asymmetries}
\title{Observations and Modeling of Line Asymmetries in Chromospheric Flares}
\author{Arkadiusz Berlicki}
\affil{Astronomical Institute, Wroc{\l}aw University, Wroc{\l}aw, Poland}
\setcounter{page}{387}

\begin{abstract}
For many years various asymmetrical profiles of different 
spectral lines emitted from solar flares have been frequently 
observed. These asymmetries or line shifts are caused predominantly 
by vertical mass motions in flaring layers and they provide a 
good diagnostics for plasma flows during solar flares.
There are many controversial results of observations and 
theoretical analysis of plasma flows in solar chromospheric 
flares. The main difficulty is the interpretation of line shifts 
or asymmetries. For many years, methods based on bisector 
techniques were used but they give a reliable results 
only for some specific conditions and in most cases cannot be 
applied. The most promising approach is to use the non-LTE 
techniques applied for flaring atmosphere. The calculation of 
synthetic line profiles is performed with the radiative transfer 
techniques and the assumed physical conditions correspond to 
flaring atmosphere. I will present an overview of different 
observations and interpretations of line asymmetries in 
chromospheric flares. I will explain what we have learnt about 
the chromospheric evaporation in the frame of hydrodynamical 
models as well as reconnection models. A critical review will 
be done on the classical methods used to derive Doppler-shifts 
for optically thick chomospheric lines. 
In particular, details on the new approach for 
interpreting chromospheric line asymmetries based on the 
non-LTE techniques will be presented. 
\end{abstract}

\section{Introduction}
Spectroscopic observations of solar chromospheric flares show that
the line profiles emitted by the flaring plasma almost always exhibit
asymmetries or shifts. These features are surely due to the chromospheric 
plasma motion and the resulting Doppler-shifts effects. Interpretation
of the shape of line profiles allows us to understand the nature of 
plasma flows during solar flares. Plasma flows in the chromosphere are important in the 
analysis of dynamics and energetics of solar flares as well as these 
phenomena supply matter to the coronal parts of flares in the process of
chromospheric evaporation. Understanding of the mechanisms which generate
flows is necessary for complete description of solar flares. 

Spectral line asymmetries arise only because of the line-of-sight component of plasma
velocity which due to the Doppler-shift is responsible for the modification
of spectral line profiles. Therefore, for the flares located on the solar 
disk center we are able to analyze flows oriented perpendicular to the 
solar surface. For simplicity, the theoretical analysis of line asymmetries
often assume that the emitting region is located on the solar disk center
and only vertical flows are present.

\begin{figure}
  \centering
  \mbox{
  \includegraphics[width=6cm]{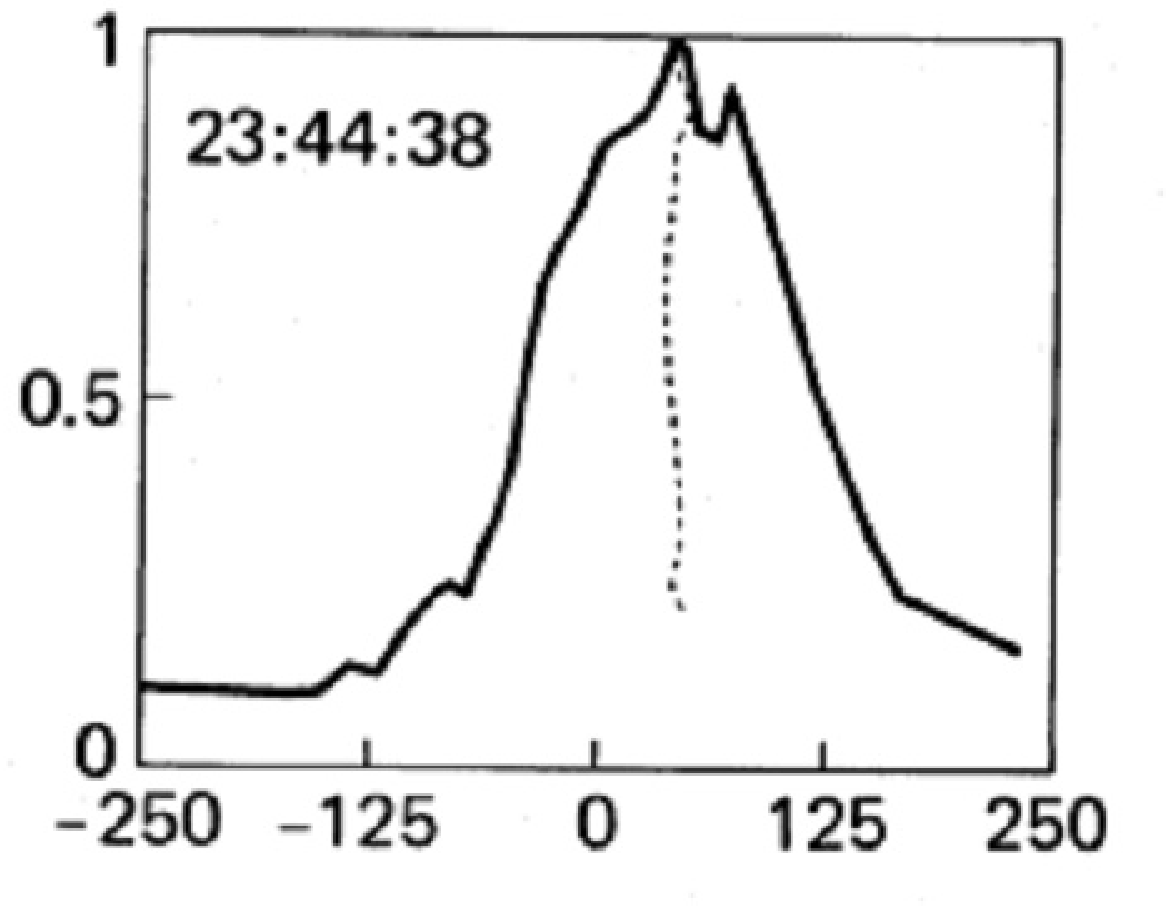}
  \includegraphics[width=5cm]{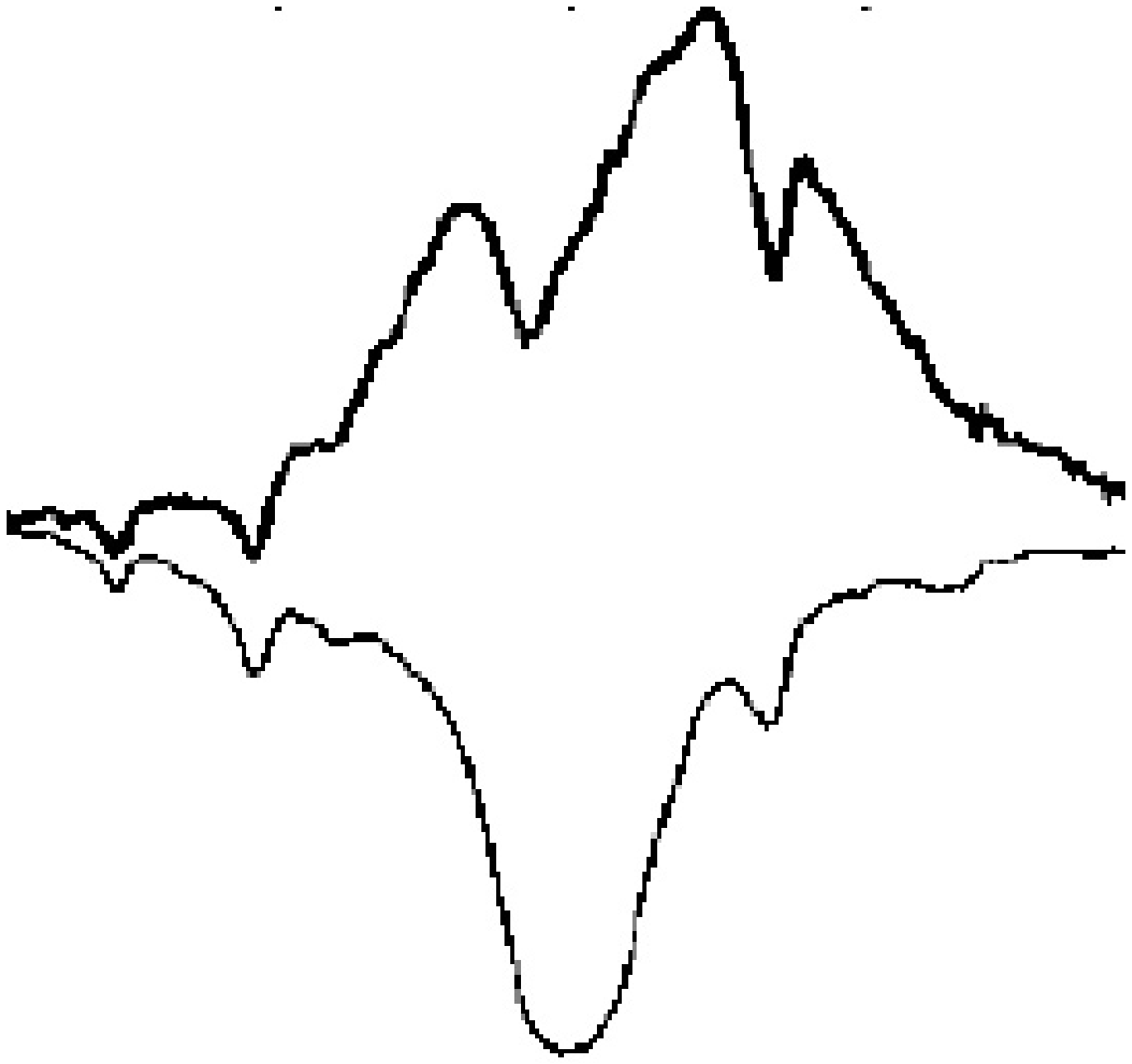}
  }
  \caption[]{Two examples of the H$\alpha$ line profiles observed
           during solar flares. {\em Left:} shifted profile -- the
           Doppler velocity may be determined from the shift of the
           whole line.  {\em Right:} complicated asymmetric profile --
           it is impossible to determine the Doppler velocity from
           such line profile using bisector methods.  }
           \label{ab-fig:prof}
\end{figure}

Unfortunately, the interpretation of the shape of line profiles is not
trivial and to determine the velocity field we must use advanced 
methods based on hydrodynamical and radiative transfer techniques.
There are some rare cases when the interpretation of line asymmetries
is relatively simple. When the flow of the whole flaring region
is homogeneous then the shape of the line is not disturbed compared to
the static symmetric line profile, but the whole line is just shifted towards
longer or shorter wavelengths (Fig.~\ref{ab-fig:prof} -- left). We can measure this shift and using
a simple Doppler formula calculate specific value of the line-of-sight velocity.
Unfortunately, in most cases the line profiles of solar flares exhibit much 
more complicated structure (Fig.~\ref{ab-fig:prof} -- right) which suggests that the velocity field is
not homogeneous and different parts of the flare move in different way.

There are two main approaches to the modeling of chromospheric line asymmetries.
One is based on the hydrodynamics calculations, where the time evolution of
the solar flare atmosphere is calculated and the radiative transfer 
formulae are used to calculate the radiation from this evolving atmosphere.
Another approach is based on semiempirical models of solar flares.

In this paper I will present a short review on observation and interpretation
of the chromospheric line asymmetries observed during solar flares. The 
term chromospheric lines is commonly used to describe all spectral lines
formed in the solar chromosphere, where the temperature is around $10^4$\,K.
These lines are formed in strong non-LTE conditions and complicated radiative
transfer calculations are necessary to describe the formation of these lines.
Strong chromospheric lines are usually optically thick what means
that the optical thickness of the plasma in these lines is very large ($\tau >> 1$).

\begin{figure}
  \centering
  \includegraphics[width=8cm]{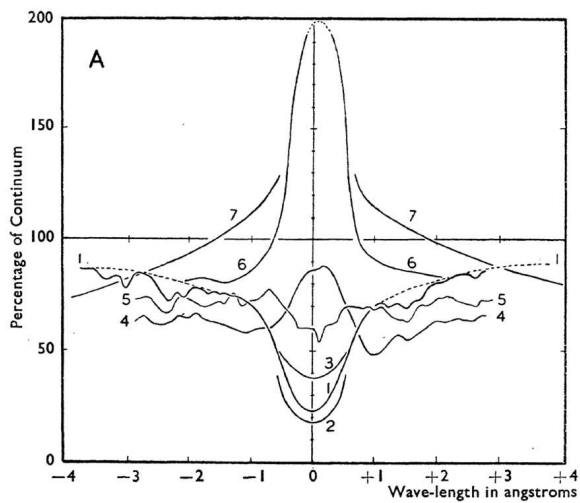}
  \caption[]{Examples of the observed H$\alpha$ asymmetric line profiles.
           Profiles 4, 5, and 6 correspond to the flare 
           emission (\cite{ab-1949MNRAS.109....3E}).
          }
  \label{ab-fig:ellison-prof}
\end{figure}

There is a wide 
literature concerning this topic and I provide the readers some 
references contained the most important results. I will concentrate
on chromospheric parts of solar flares where the emission comes from cool ($10^4$\,K)
plasma. This emission is produced mostly in strong resonance lines of 
hydrogen, calcium or magnesium (H$\alpha$, H$\beta$, H$\delta$, Ca\,II\,H, K, etc.) 
Since the most spectroscopic observations performed during past years 
concern H$\alpha$ line, the interpretation of these results will take 
considerable part of this review.

\section{Early Observations and Interpretations}

Asymmetries of chromospheric lines emitted by the flaring plasma have been 
observed for more than half of the century. After the solar spectrographs
were developed in order to produce the solar spectrum with sufficient
spectral resolution, observers noticed that the line profiles emitted during
solar flares are not symmetric (Fig.~\ref{ab-fig:ellison-prof}). 
It was clear from the beginning that these kinds of modifications
of the spectral lines are due to the mass motion driven during solar flares.
However, the mechanisms which could drive the plasma flows were not
known yet at those days.

\begin{figure}
  \centering
  \includegraphics[width=10cm]{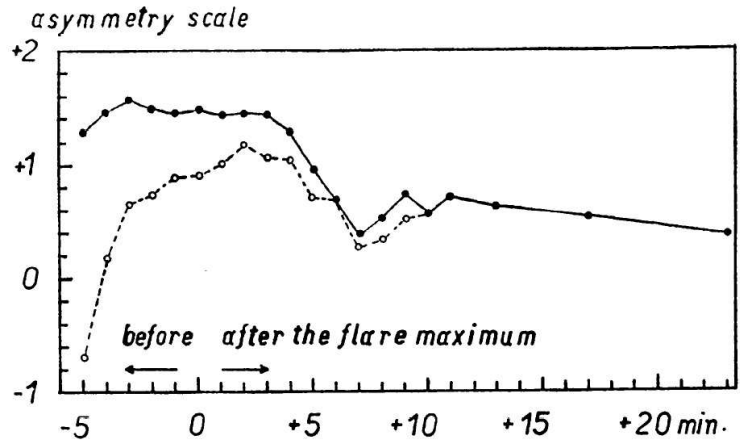}
  \caption[]{Absolute mean value of the asymmetry of the H$\alpha$ line profiles
           as a function of time development of flares (solid line). 
           Dashed curve takes positive and negative signs into account. Time of flare
           maximum = 0 (\cite{ab-1962BAICz..13...37S}).
          }
  \label{ab-fig:asymm-evol}
\end{figure}

The first analysis of the line asymmetries were concentrated on statistical
description of the behaviour of lines. \citet{ab-1962BAICz..13...37S} presented 
a qualitative analysis of 244 H$\alpha$ and Ca\,II\,K spectra of 92 flares. They found that
the blue asymmetry (blue wing enhancement) occurs mainly in the early phase
of flares, before flare maximum. However, only 23\% of flares contain at 
least one region with blue asymmetry. 80\% of flares exhibit the red asymmetry 
which dominates during and after the maximum of flare. It is worth to 
notice that only 5\% of flares shows blue asymmetry exclusively. However, because
not all flares were observed from their beginning, the occurrence of blue asymmetry
may be missed for many flares. In Fig.~\ref{ab-fig:asymm-evol}
the time evolution of the asymmetry is presented. Similar analysis was
performed by \citet{ab-1983SoPh...83...15T}. By inspection of 
off-band filtergrams of 60 flares obtained
in $\pm$1 and $\pm$2\,\AA\ from the H$\alpha$ line center he found that 92\%
of flares show red asymmetry and only 5\% show blue asymmetry.

Early statistical analysis gave us an qualitative idea about the line
asymmetry but the physical interpretation of the shape of line 
profiles is not considered. The asymmetric line profile contains
an information about the velocity field in the region where the observed
line is formed. Therefore, the first problem which needs to be solved is 
to determine this velocity field using the observed line profile. Secondly,
we have to answer the question why do we observe such a flow of plasma, 
what can generate the flows? 

Determination of the velocity from the observed profiles of the 
chromospheric lines is not a trivial task due to the complicated
processes of line formation and complex velocity field in the 
chromosphere. Strong chromospheric lines are optically thick and 
the radiation observed at different parts
of the line profile comes from different height $z$ across 
the chromosphere (\cite{ab-1978stat.book.....M}).
The core of strong chromospheric lines (hydrogen Balmer lines, Ca lines)
is formed much higher than the wings of these lines.
Therefore, the asymmetry of the specific line depends on the relation
between the height $z$ of formation of given part of the line profile
and the value of the velocity at this height. This means that if
the function describing the velocity across the height in the chromosphere
is complicated, the emergent line profile has also very complicated shape.

Despite of all difficulties with determining correctly the velocity from
the line profile shape, many authors tried to use H$\alpha$, Ca\,II and other
lines to find the velocity in the flaring chromosphere. All these 
determinations were based on the measurements of Doppler-shifts of the line 
cores or, more commonly, on the bisector technique. 

Using the shift of the line core to obtain the velocity may be 
misleading and the determined velocity is not correct when
the velocity gradient in the chromosphere is significant (\cite{ab-1970SoPh...12..175A}). 
Unfortunately, the estimation of the Doppler-shift obtained with the 
bisector technique can also give wrong results. The Doppler-shift 
of the line profile correspond to the shift of the central point of 
the bisector connecting the two wings of the spectral line.
Since the different parts of the spectral
line are formed at different height in the chromosphere, using bisectors
connecting the wings observed at different frequency, we can estimate the 
line-of-sight velocity at different layers of the chromosphere. However, because
the radiation of specific frequency within the spectral line does not come 
from one narrow layer of the chromosphere but rather from geometrically thick
region, we cannot say that the Doppler velocity determined form given
bisector correspond to plasma flow at given height in the chromosphere.
Moreover, if the velocity gradient in the chromosphere is large then the
bisector method cannot be used because the Doppler-shift of any bisector
results from a superposition of many shifts due to the motion of plasma 
with different velocity along line-of-sight.  
In spite of this the bisector technique was commonly used for many years
until it was replaced by more advanced, complicated but much more
precise non-LTE radiative transfer techniques with velocity field included.

\begin{figure}
  \centering
  \includegraphics[width=12cm]{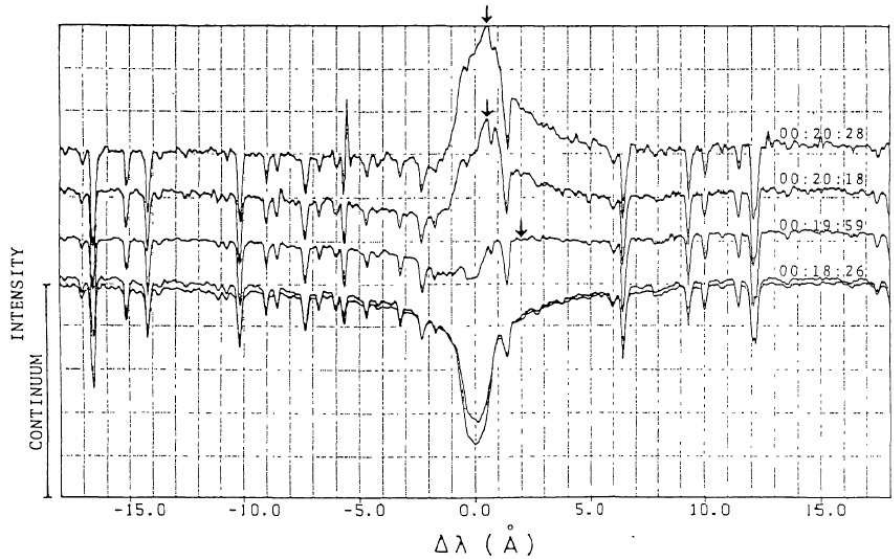}
  \caption[]{Temporal variation of the H$\alpha$ line profiles observed during
           solar flare. The black arrow indicates the peak of asymmetric lines. The 
           bottom profile represents the emission of the quiet-Sun 
           area (\cite{ab-1984SoPh...93..105I}).
          }
  \label{ab-fig:ichim-prof}
\end{figure}

One of the first interpretation of chromospheric line asymmetry observed 
during flare was presented by \citet{ab-1982ApJ...263..409A}. The authors postulated that 
these asymmetries are due to chromospheric evaporation driven
by accelerated electrons or thermal conduction. 
\citet{ab-1968ApJ...153L..59N} was probably the first who realise that chromospheric plasma 
heated during solar flare may evaporate. This evaporated plasma provides
material for loop prominences often observed as the
so-called post-flare loops (\cite{ab-1976SoPh...50...85K}; \cite{ab-1978ApJ...220.1137A}). For more 
complete review of chromospheric evaporation see Hudson paper in this book.
\citet{ab-1982ApJ...263..409A} postulated that in the analysed flare the non-thermal electrons
heat the chromosphere mainly during the impulsive phase,
while thermal conduction from the hot coronal plasma heated earlier 
dominates during the late, thermal phase. Both mechanisms drive upflow of the 
cool plasma. The authors stress that for the first
time they observed chromospheric evaporation in H$\alpha$ line.

\begin{figure}
  \centering
  \includegraphics[width=6cm]{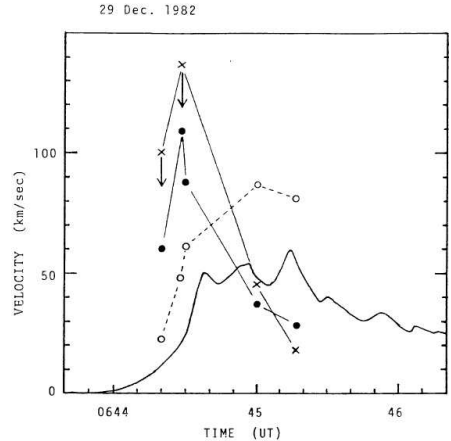}
  \caption[]{An example of temporal variation of the downflow velocity observed 
  in flaring region. Filled circles correspond to the values obtained from
  the shift of the H$\alpha$ far wings, while crosses to the values
  obtained from the shift of the line peak. Open circles represent the
  time evolution of the H$\alpha$ intensity. The microwave emission
  at 3750 MHz is also shown (\cite{ab-1984SoPh...93..105I}).
          }
  \label{ab-fig:veloc-evol}
\end{figure}

Contradictory results were published by \citet{ab-1984SoPh...93..105I}
who suggested that during the impulsive phase of solar flares the
downflow of the cool chromospheric plasma is present. These results
are based on the large red asymmetry of the H$\alpha$ line observed
during the impulsive phase of many solar flares
(Fig.~\ref{ab-fig:ichim-prof}).  The downward motion increases at the
onset of a flare to its maximum velocity of \mbox{40 to 100
$\mathrm{km\;s^{-1}}$} shortly before the impulsive peak of microwave
emission, and rapidly decreases before the H$\alpha$ reaches its
maximum (Fig.~\ref{ab-fig:veloc-evol}). The red asymmetry of the
H$\alpha$ line may be also explained by the attenuation of the blue
wing by the rising plasma over the flare but the authors exclude this
case because the optical thickness of a rising cloud is too small to
explain the emission deficit of the blue wing of H$\alpha$ line. Also
the high temporal resolution spectroscopic observations of H$\alpha$
line performed by \citet{ab-1987SoPh..114..115W} confirm the existence
of red asymmetry during the impulsive phase of solar flares. The
largest asymmetry is observed during the maximum of microwave emission
what confirms the earlier results that the plasma downflow is driven
by the accelerated non-thermal electrons.

\begin{figure}
  \centering
  \includegraphics[width=10cm]{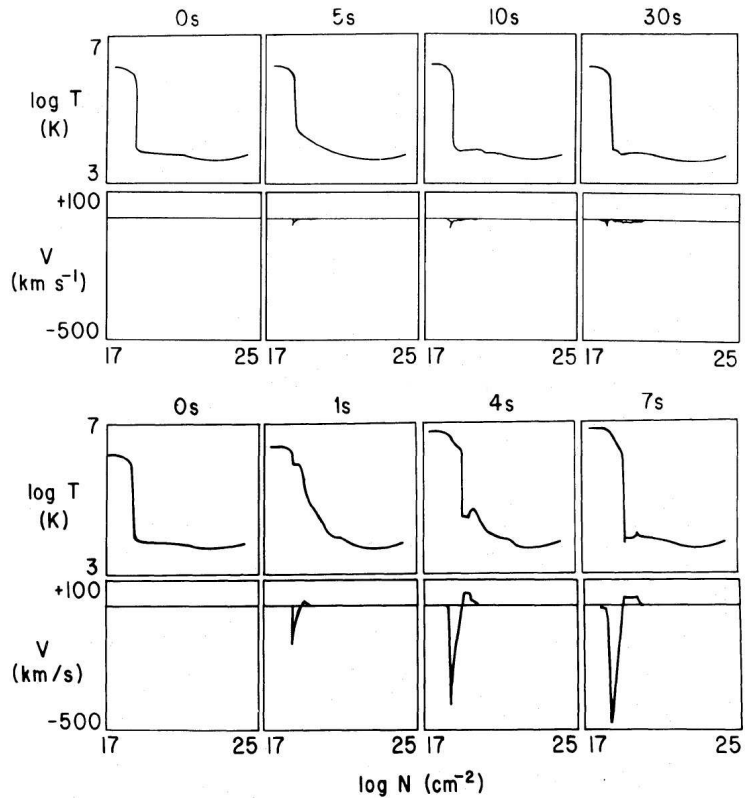}
  \caption[]{Time evolution of the temperature and velocity in the 
   loop atmosphere heated by the low (upper two panels) and the high (lower
   two panels) flux of non-thermal electrons. In the first case the upflow
   with the low velocity in the transition region and in the chromosphere is 
   present (gentle evaporation). For the strong flux of non-thermal electrons
   the high-velocity upflow (up to $-500$\,km\,s$^{-1}$) in the high temperature
   region is obtained (explosive evaporation) and downflow 
   (up to $+40$\,km\,s$^{-1}$) is observed
   in the chromosphere (chromospheric condensations). For more details see
   the paper of \citet{ab-1985ApJ...289..414F}.
          }
  \label{ab-fig:fisher-1}
\end{figure}

\section{Hydrodynamic Modeling of the Flows}

The downflow of cool plasma in the form of chromospheric condensations
observed during solar flares was predicted theoretically by
\citet{ab-1985ApJ...289..414F}.  If a region of the chromosphere
heated by non-thermal electrons is thick enough, then the rapid
temperature increase produces an enhanced pressure in the heated
region. This overpressure, besides the evaporation, also drives
downward moving cool and dense chromospheric condensations
(\cite{ab-1985ApJ...289..434F}) which seem to be responsible for red
asymmetry of the H$\alpha$ line profiles reported by many authors.
\citet{ab-1985ApJ...289..434F} modeled the hydrodynamic and radiative
response of the atmosphere to short impulsive injections of
non-thermal electron beams (Fig.~\ref{ab-fig:fisher-1}). They showed
that a high-energy flux of non-thermal electrons drives explosive
evaporation accompanied by the formation of cool chromospheric
condensations in the flare chromosphere.  A different situation occurs
when the flux associated with non-thermal electrons is very low. Then
only a weak chromospheric evaporation takes place. This kind of
evaporation is referred to as gentle evaporation
(\cite{ab-1978ApJ...220.1137A}; \cite{ab-1987ApJ...317..956S}) and it
can be observed in chromospheric spectral lines like H$\alpha$ or in
Ca\,II~8542\,\AA.  \citet{ab-1978ApJ...220.1137A} suggested that the
gentle chromospheric evaporation may also occur after the primary
energy release, when the non-thermal electron flux is stopped. This
evaporation could be driven by the large conductive heat flux from a
high temperature flare plasma contained in magnetic tubes above the
chromosphere. Such physical conditions may appear during the gradual
phase of solar flares, when there is no significant flux of
non-thermal electrons. In the \citet{ab-1989SoPh..120..285F} model for
flare-loop formation by magnetic reconnection the conduction of the
thermal energy generated at the slow-mode shocks drives a gentle
evaporative upflow from the ribbons.

\begin{figure}
  \centering
  \includegraphics[width=8cm]{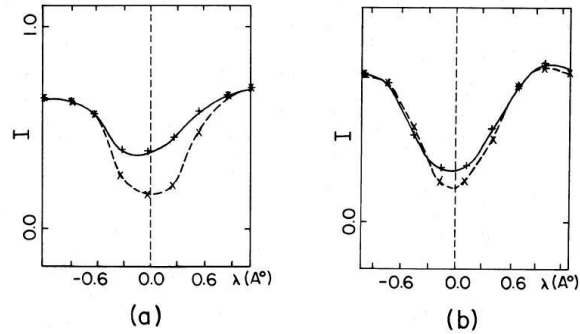}
  \caption[]{H$\alpha$ line profiles observed during the gradual phase
  of a solar flare (solid lines). Weak blue-shift of these lines
  suggests slight upflow of the plasma with the velocity of a few km\,s$^{-1}$
  interpreted as gentle evaporation. Dashed lines represent the reference 
  line profiles of the quiet Sun area (\cite{ab-1987ApJ...317..956S}).
          }
  \label{ab-fig:schmieder-gentle}
\end{figure}

\citet{ab-1987ApJ...317..956S} observed small but long-lasting blue-shifts in flare
ribbons in the H$\alpha$ line during the gradual phase of three solar
flares and interpreted them as due to upflows with velocities
less than 10\,km\,s$^{-1}$ (Fig.~\ref{ab-fig:schmieder-gentle}). These upflows were 
believed to be caused by gentle
chromospheric evaporation driven by the heat conduction
along the field lines connecting the chromosphere with a reconnection
site in the corona.

\begin{figure}
  \centering
  \mbox{
  \includegraphics[width=6cm]{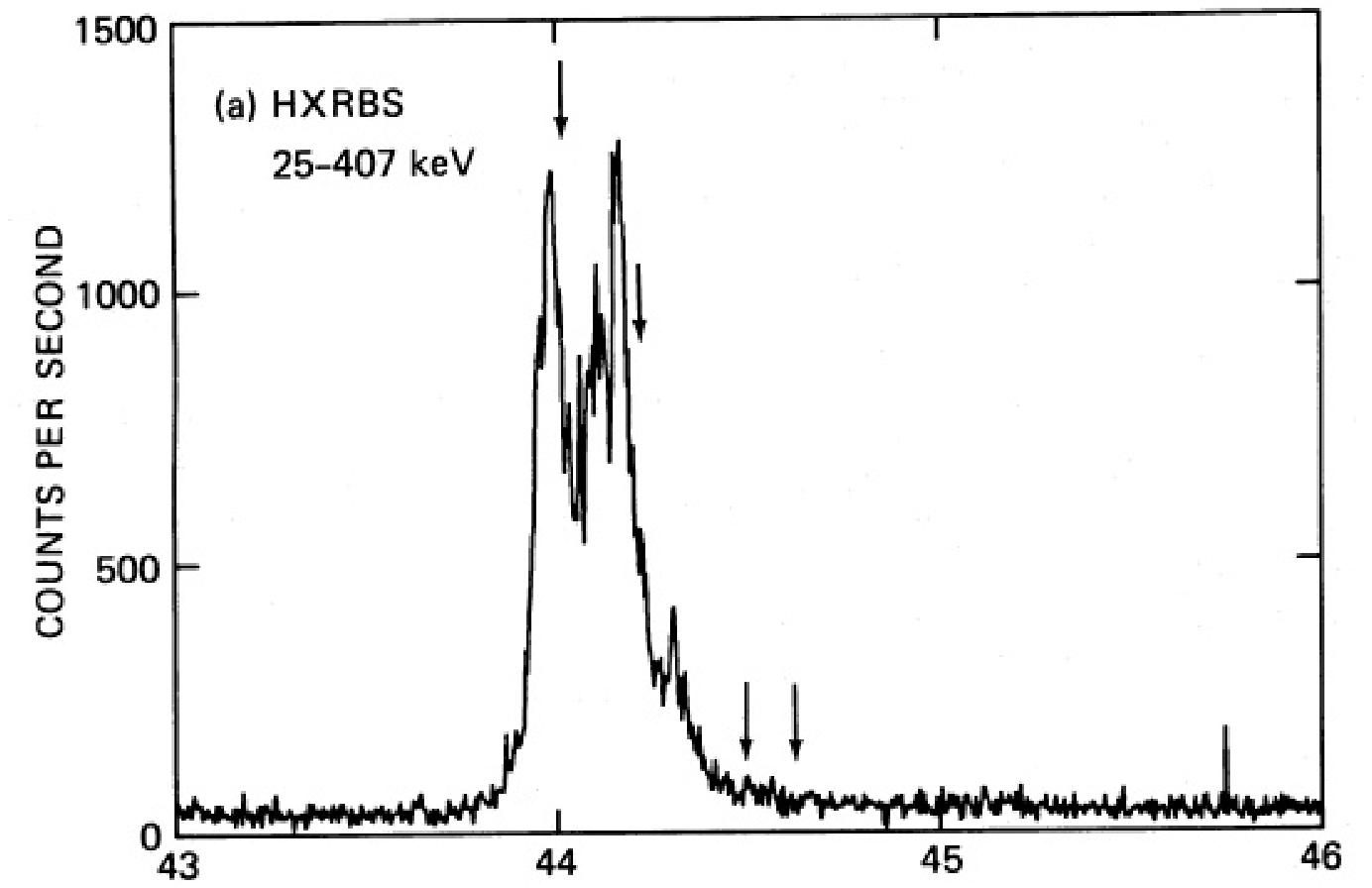}
  \includegraphics[width=5.5cm]{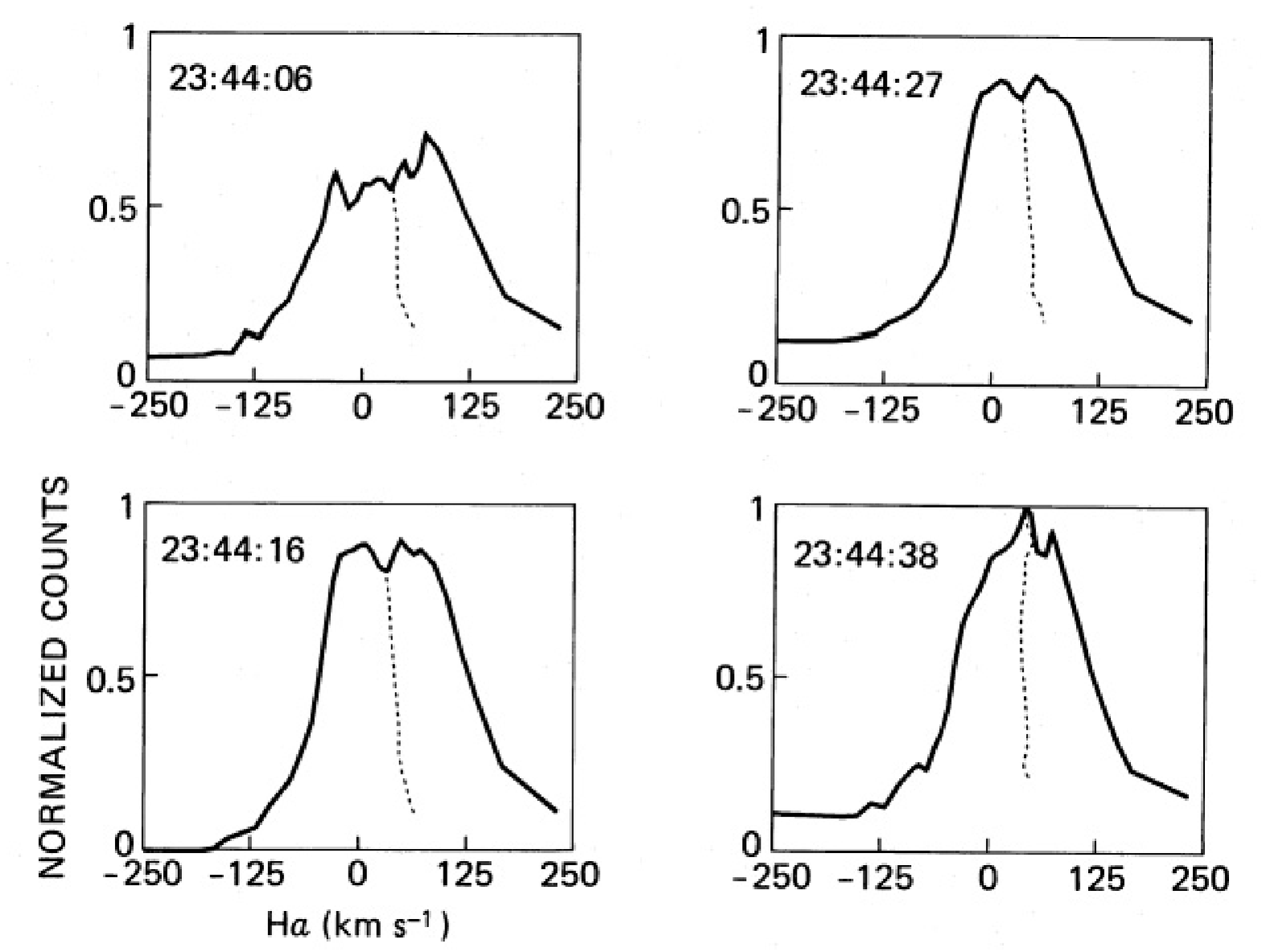}
  }
  \caption[]{The hard X-ray emission (left) and the H$\alpha$ excess line profiles 
  (right) observed during the impulsive phase of a solar flare. Dashed lines mark 
  the centroids of the profile. The obtained downward velocities are around  
  50\,km\,s$^{-1}$. The numbers on the X-axis of left panel represent
  minutes after 23 UT and the arrows -- the times of H$\alpha$ observed
  line profiles (\cite{ab-1988ApJ...324..582Z}).
          }
  \label{ab-fig:zarro-1}
\end{figure}

\begin{figure}
  \centering
  \includegraphics[width=2.5cm]{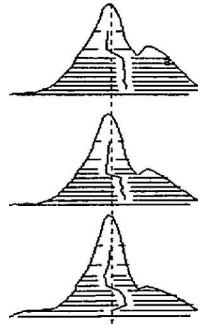}
  \caption[]{An example of the H$\alpha$ line profiles observed during the 
           impulsive phase of a solar flare. The red wing enhancement is observed
           at the same time as strong hard X-ray emission. Note that the central
           part of the line is slightly blue-shifted (\cite{ab-1989ApJ...338L..33Z}).
          }
  \label{ab-fig:zarro-2}
\end{figure}

The downflow of cool chromospheric plasma during the impulsive phase of
solar flares predicted in the theoretical calculations 
was reported by many authors. \citet{ab-1988ApJ...324..582Z} observed large red asymmetry
of the H$\alpha$ line during the period of hard X-ray burst (Fig.~\ref{ab-fig:zarro-1}). 
These asymmetries were used to determine the downward velocities estimated
from the maximum shift of the centroid of the bisectors. The averaged 
over all red-shifted pixels during the impulsive phase velocity was of
the order of $60 \pm 10$\,km\,s$^{-1}$. The downflow analysed in the 
H$\alpha$ data and the upflow observed in the X-ray lines allows the 
authors to analyze the momentum balance of the flow. They conclude that the
momenta of upflowing and downflowing plasma are approximately equal. 
\citet{ab-1989ApJ...338L..33Z} conclude that the downflow velocity measured
from red wing enhancement can be used as a diagnostics of impulsive solar
flare heating conditions (Fig.~\ref{ab-fig:zarro-2}).

The work of \citet{ab-1995SoPh..158...81D} shows that the velocity of chromospheric 
downflows deduced from the red asymmetry of H$\alpha$ line is around 
30~--~40\,km\,s$^{-1}$ with the lifetime of the
order of 2~--~3 minutes. There are two major problems found by the
authors: Why is the line center nearly not shifted while the line
wing shows great asymmetries? The second problem concerns the life
time of the downflow which is considerably longer than the life
time predicted in \citet{ab-1989ApJ...346.1019F} simulations. Recent hydrodynamic
and radiative transfer simulations may now explain these two problems.

\begin{figure}
  \centering
  \includegraphics[width=9cm]{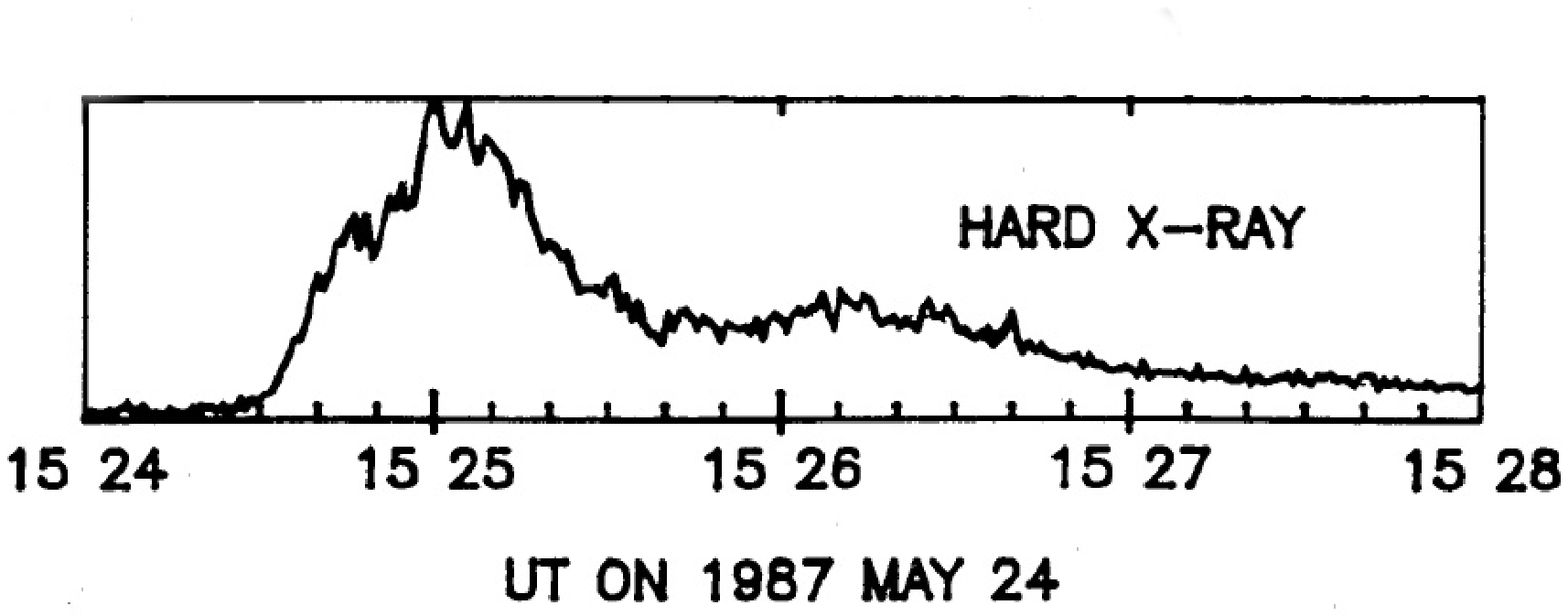} \\
  \includegraphics[width=4cm]{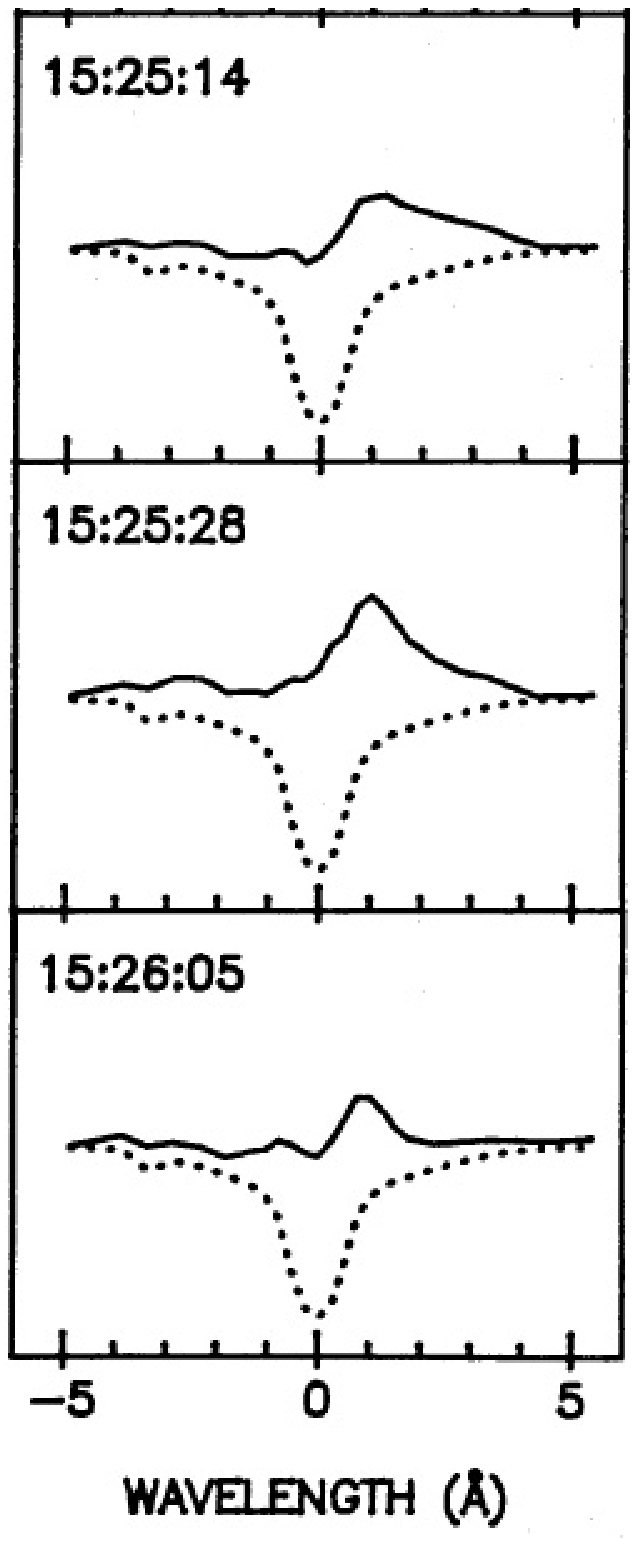}
  \caption[]{The hard X-ray emission (top) and the H$\alpha$ line profiles 
           with the red asymmetry (bottom) observed during the impulsive phase 
           of a solar flare. Dotted lines mark 
           the reference quiet Sun profile (\cite{ab-1989ApJ...341.1088W}).
          }
  \label{ab-fig:wulser}
\end{figure}

There are more papers presenting the observations of the red asymmetry
related to the chromospheric condensations driven during the impulsive phase
of solar flares (e.g., \citet{ab-1989ApJ...341.1088W} -- Fig.~\ref{ab-fig:wulser}). 
All of them determine more or less consistent observational 
picture of the chromospheric flows during the impulsive phase of flares.
This picture is based on many spectroscopic observations of the chromospheric
line profiles. An important step was done when it became possible to
calculate theoretical line profiles and compare them with observations.

\begin{figure}
  \centering
  \includegraphics[width=10cm]{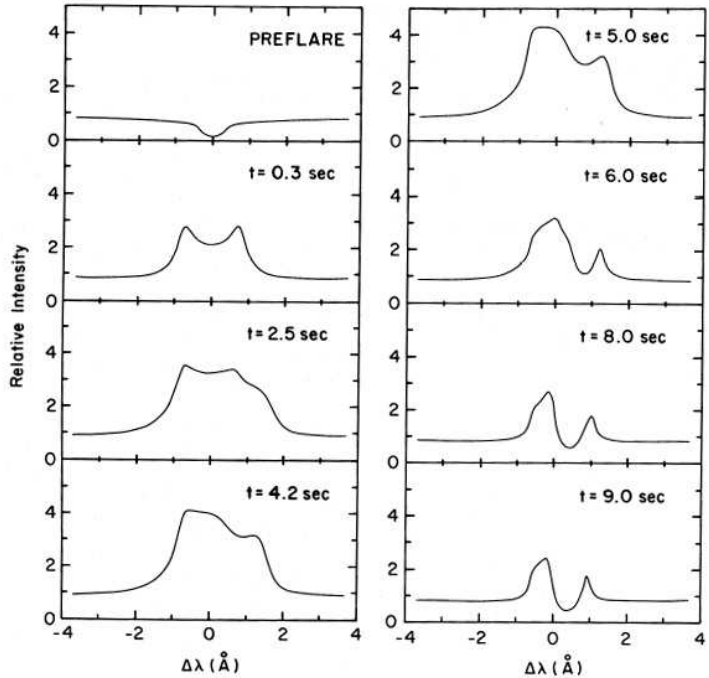}
  \caption[]{Time evolution of the calculated H$\alpha$ line profiles for the 
          electron beam heated model atmosphere 
          of \citet{ab-1985ApJ...289..414F} (\cite{ab-1987ApJ...322..999C}).
          }
  \label{ab-fig:canf-gal}
\end{figure}

\citet{ab-1987ApJ...322..999C} computed time-dependent H$\alpha$ line
profiles for the dynamic model atmosphere of \citet{ab-1985ApJ...289..414F}. They simulate
the effects of power-law electron beam heated chromosphere. Solving the
radiative transfer equations for one-dimensional model atmosphere the evolution of H$\alpha$
line profile was estimated. The time of the electron beam heating was
5 s and for detailed description of other parameters and computational
methods see \citet{ab-1987ApJ...322..999C}. In Fig.~\ref{ab-fig:canf-gal}
the time sequence of the H$\alpha$ line profile is presented.
During the non-thermal heating, the red-shifted component is present but after the 
heating was stopped, the H$\alpha$ line exhibit the blue asymmetry although
the central absorption feature is shifted towards longer wavelengths.
This behaviour is explained by downflow of the chromospheric
condensation. It is also worth to notice that the response of the H$\alpha$ 
emission to the non-thermal electron beam is very fast (less than second).

Similar but more precise simulations of the dynamics and radiation in a solar 
flare loop was presented by \citet{ab-1999ApJ...521..906A}. Except the non-thermal
heating of the chromosphere, they took into account the thermal heating
by the soft X-ray irradiation within 1~--~250\,\AA\ range. \citet{ab-1997ApJ...481..500C} 
radiative-hydrodynamic code was used to analyze the response of the lower 
atmosphere at the footpoint of a flare loop. In the radiative transfer calculations 
the important transitions of hydrogen, helium and singly ionized calcium and
magnesium were treated in non-LTE. One-dimensional atmospheric model
was used in the calculations. 

\begin{figure}
  \centering
  \includegraphics[width=7cm]{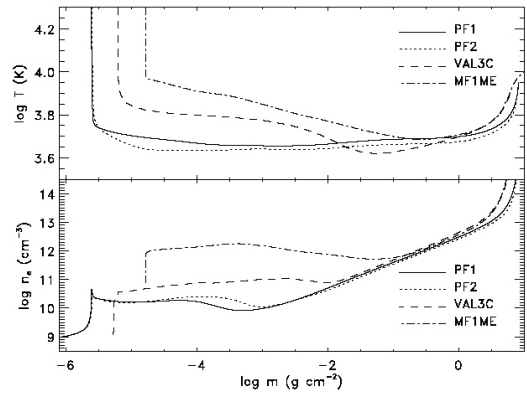}
  \caption[]{The temperature and electron density stratifications of both 
           preflare atmospheres PF1 and PF2 compared to the standard semiempirical 
           VAL3C chromospheric model of \citet{ab-1981ApJS...45..635V} and to 
           the semiempirical active atmosphere MF1ME of \citet{ab-1990PhDT.........6M}
           (\cite{ab-1999ApJ...521..906A}).
           }
  \label{ab-fig:abbett-hawley-1}
\end{figure}

As a starting models the authors took two different cases PF1 and PF2. 
The temperature and electron density stratifications of both preflare atmospheres 
are shown in Fig.~\ref{ab-fig:abbett-hawley-1}. Three levels of the non-thermal
heating was considered which correspond to weak (F9), moderate (F10), 
and strong (F11) non-thermal flare heating. The PF1 atmosphere is heated
for 70 s with the F9 and F10 fluxes, and the PF2 atmosphere is
heated for a shorter, 4 s burst but with strong F11 heating.

\begin{figure}
  \centering
  \mbox{
  \includegraphics[width=6cm]{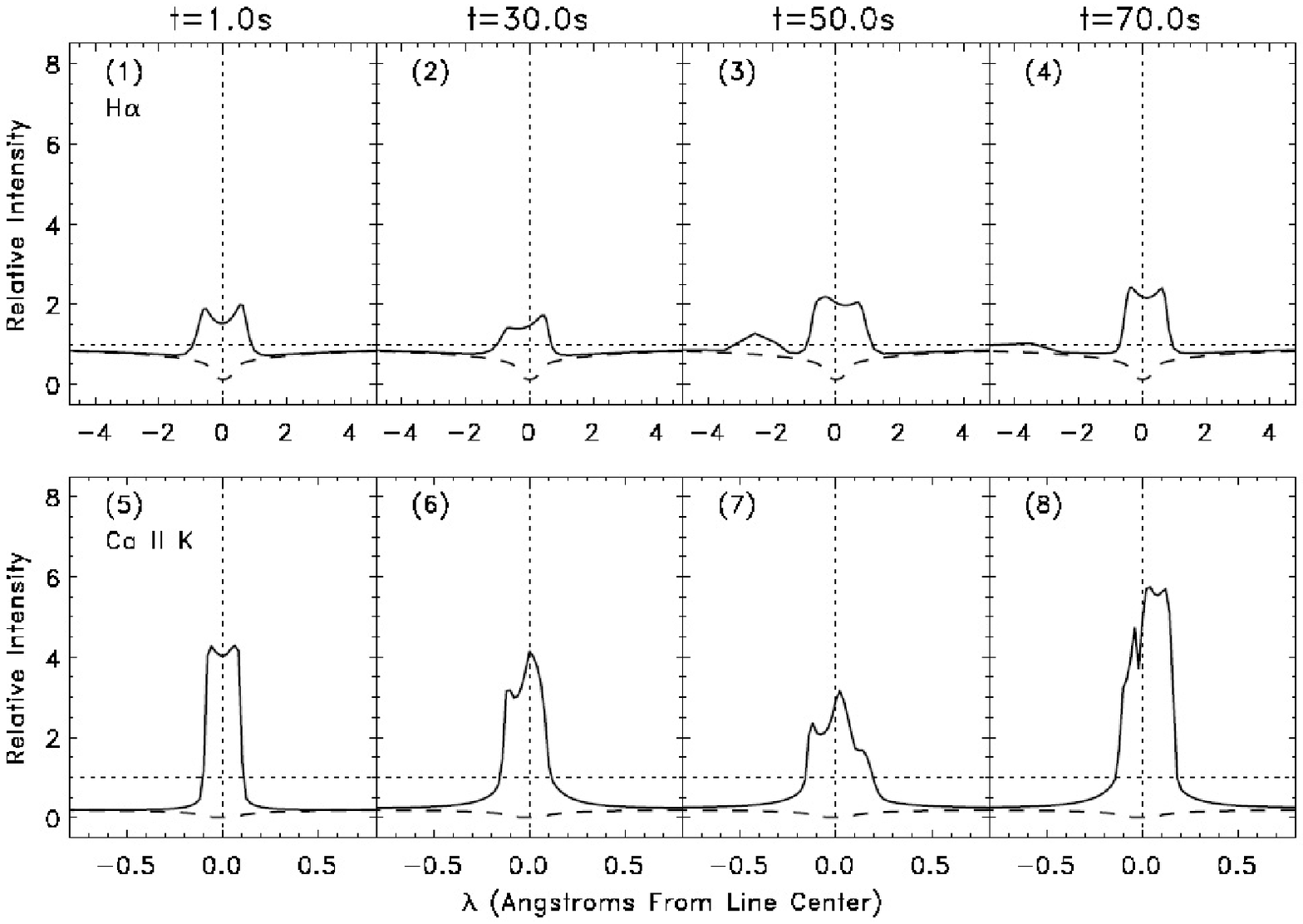}
  \includegraphics[width=6cm]{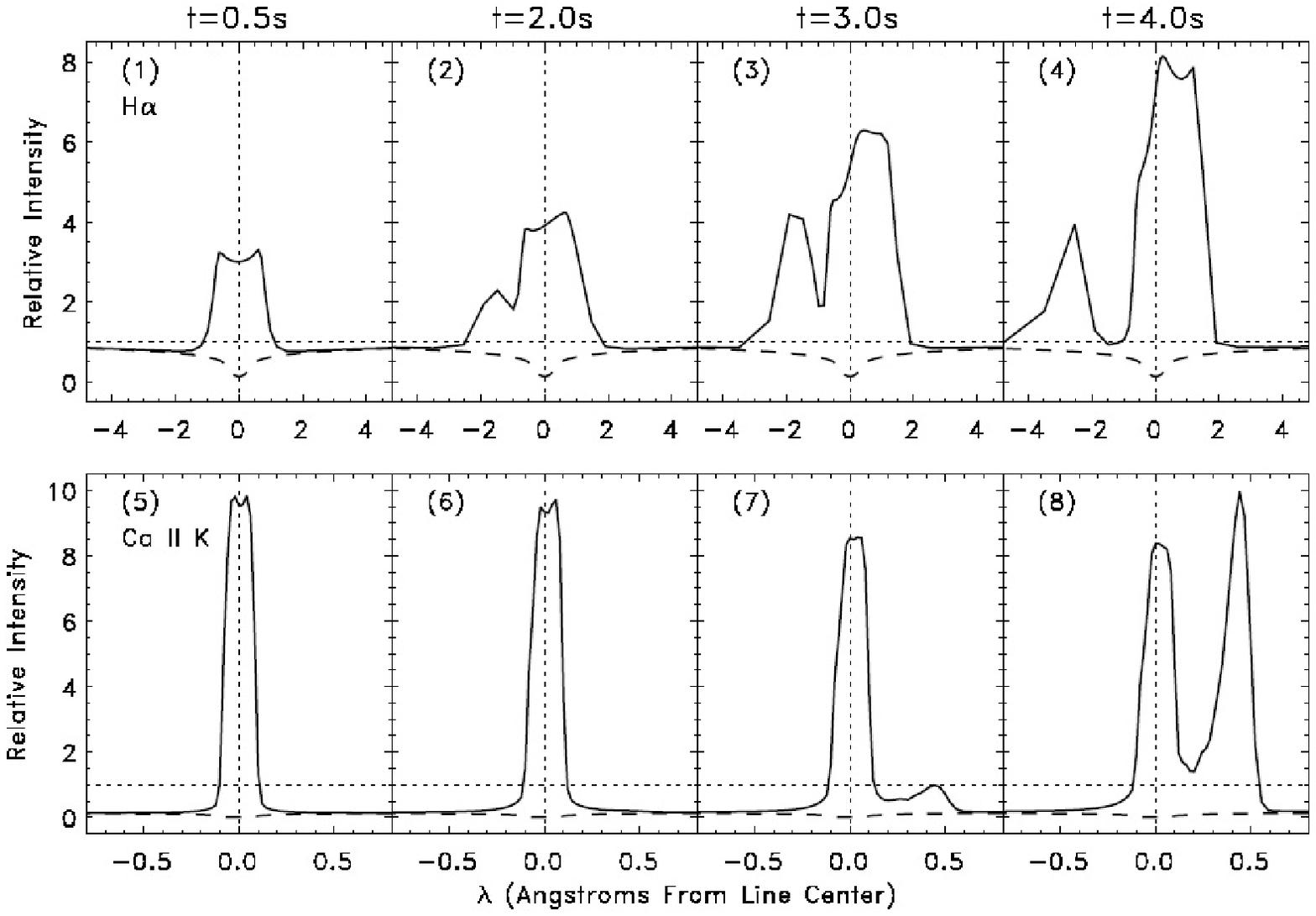}
  }
  \caption[]{{\em Left:} Time evolution of the H$\alpha$ and Ca\,II\,K line profiles 
           for moderate heating of the atmosphere by non-thermal electrons 
           (model F10). In each panel, the vertical axis represents relative 
           intensity with respect to the continuum level and the horizontal 
           axis denotes the wavelength from the line centre. The dashed lines 
           in each panel represent the preflare line profile. {\em Right:} Time 
           evolution of the H$\alpha$ and Ca\,II\,K line profiles but calculated
           for strongly heated atmosphere (model F11) (\cite{ab-1999ApJ...521..906A}).
           }
  \label{ab-fig:abbett-hawley-2}
\end{figure}

\begin{figure}
  \centering
  \includegraphics[width=12cm]{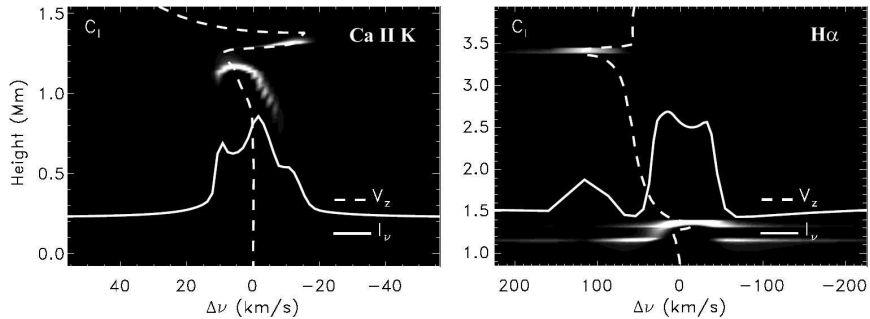}
  \caption[]{The approximate formation height of different parts of the spectral line profile
           may be described by contribution function. This figure presents components 
           of the intensity contribution function for Ca\,II\,K (left) and H$\alpha$ (right)
           lines after 50 s of flare heating in model F10. Line frequencies are in velocity 
           units. Atmospheric velocities (dashed lines) are taken to be positive toward the 
           corona. Thus, negative velocities associated with downward-moving material 
           correspond to red-shifts in the profile (\cite{ab-1999ApJ...521..906A}).
          }
  \label{ab-fig:abbett-hawley-3}
\end{figure}

Figure~\ref{ab-fig:abbett-hawley-2} presents the time evolution of
emergent H$\alpha$ and Ca\,II\,K line profiles. For the H$\alpha$ line
separated blue-shifted component is clearly visible while Ca\,II line exhibit
red-shifted component. The contribution function calculated for these 
two lines explain why we observe such a two-components and
asymmetric profiles (Fig.~\ref{ab-fig:abbett-hawley-3}).

This analysis shows that the evolution of non-thermally heated
chromosphere progresses through two distinct dynamic 
phases (\cite{ab-1999ApJ...521..906A}): a gentle phase, where
the non-thermal energy input of the flare is essentially radiated
away into space, and an explosive phase, where the
flare energy rapidly heats the atmosphere and drive large
amounts of chromospheric material up into the corona, and
down toward the photosphere. During the explosive phase,
there is significant plasma motion and there are steep
velocity gradients. Moreover, the effects of
thermal X-ray heating of the chromosphere remain negligible
compared to the non-thermal heating in the impulsive phase.

\begin{figure}
  \centering
  \mbox{
  \includegraphics[width=6cm]{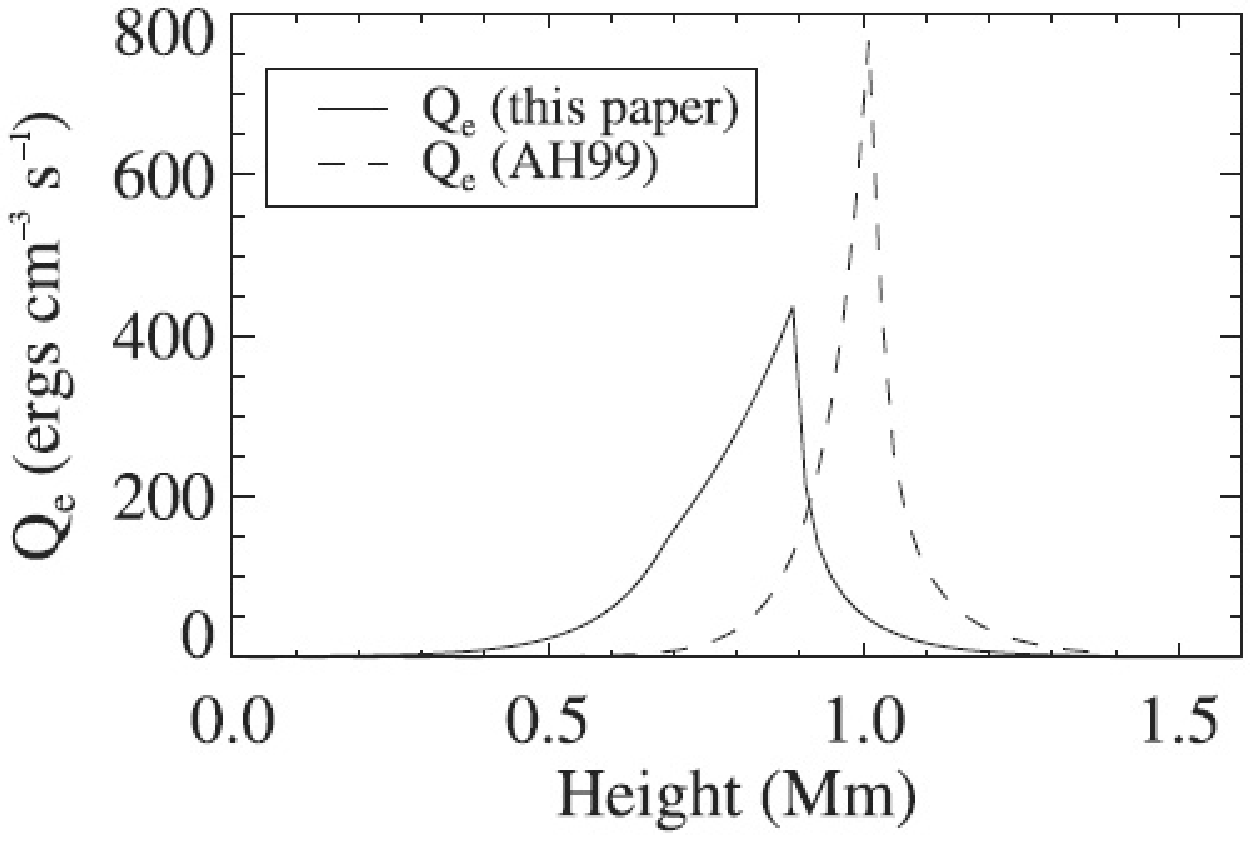}
  \includegraphics[width=6cm]{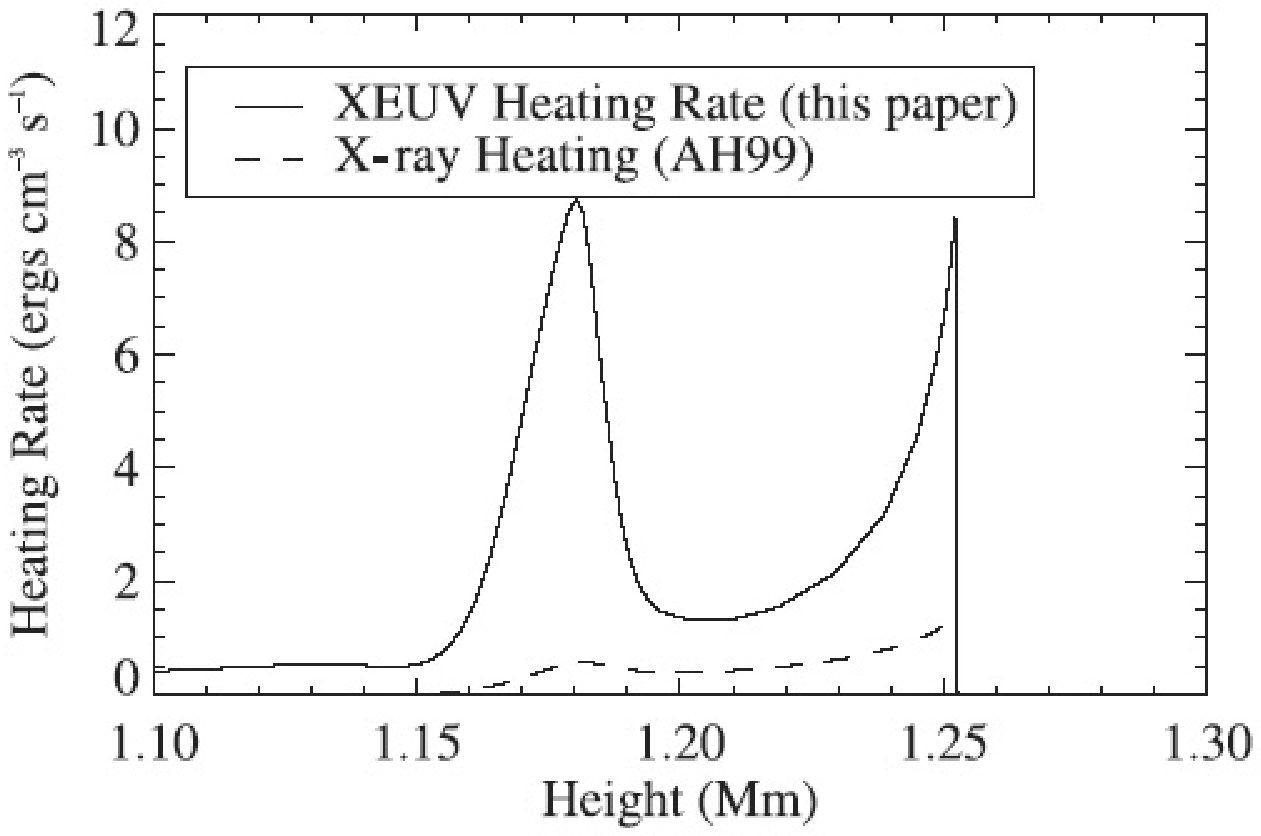}
  }
  \caption[]{{\em Left:} Comparison of the electron beam heating rate in the preflare 
           atmosphere for the F10 flare model. The solid line shows the heating 
           rate used in the paper of \citet{ab-2005ApJ...630..573A}, and the dashed line 
           shows the initial beam heating rate of \citet{ab-1999ApJ...521..906A}.
           {\em Right:} Solid line -- thermal XEUV heating used in \citet{ab-2005ApJ...630..573A}, 
           dashed line -- soft X-ray heating rate used previously by 
           \citet{ab-1999ApJ...521..906A}.
          }
  \label{ab-fig:allred-1}
\end{figure}

Similar, but much more extended calculations were presented
by \citet{ab-2005ApJ...630..573A}. The basics of computational methods
are similar to those described by \citet{ab-1999ApJ...521..906A} but there
are some significant improvements. 
The authors include the double power-law
electron beam energy distributions recently observed in solar flares
with the Reuven Ramaty High-Energy Solar Spectroscopic Imager
(RHESSI) satellite. Additionally, the effects
of XEUV heating from a large number of high-temperature
lines was taken into account using results from the CHIANTI and ATOMDB 
databases and a wide range 1~--~2500\,\AA\ was used for direct
thermal heating of the chromosphere. Figure~\ref{ab-fig:allred-1} present
the comparison of the heating rates used by \citet{ab-1999ApJ...521..906A}
and \citet{ab-2005ApJ...630..573A}.

\begin{figure}
  \centering
  \includegraphics[width=12cm]{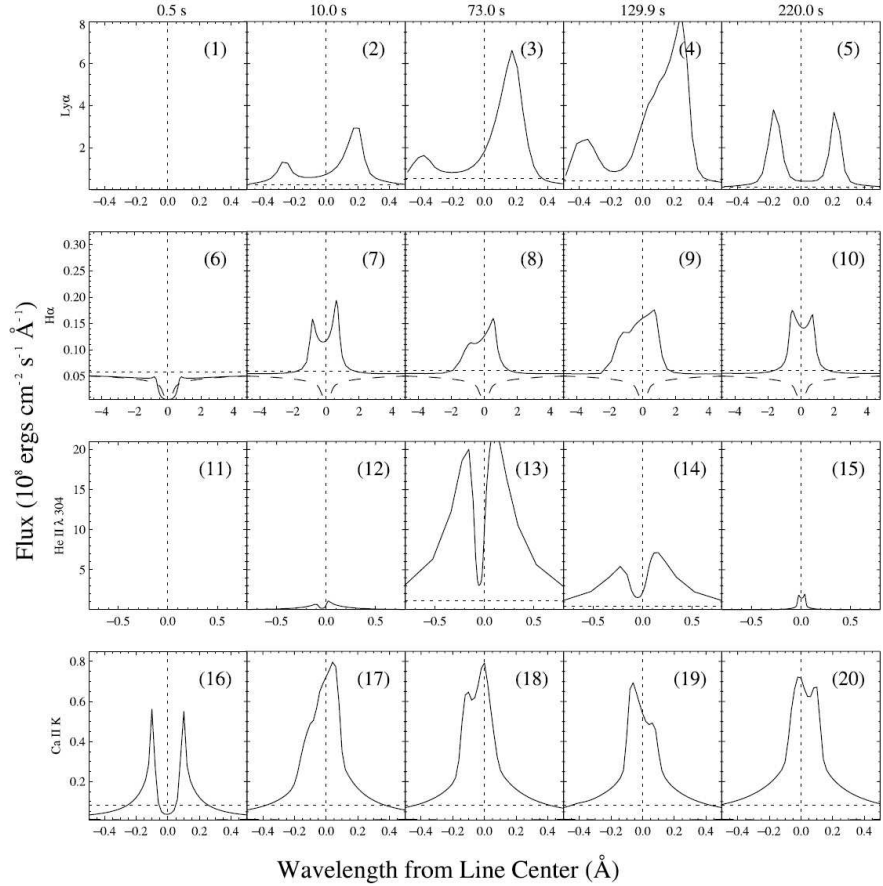}
  \caption[]{Time evolution of the synthetic line profiles of Ly-$\alpha$, H$\alpha$, 
           He\,II~304\,\AA, and Ca\,II\,K lines for the moderate level
           of non-thermal heating (model F10). Times are indicated at the 
           top of each column. The dotted lines indicate the level of the continuum 
           close to the line center, while the dashed line is the preflare line 
           profile (\cite{ab-2005ApJ...630..573A}).
  }
  \label{ab-fig:allred-2}
\end{figure}

These new calculations confirmed the previous results. However, the line
profiles evolution differs from the \citet{ab-1999ApJ...521..906A} 
calculations -- the line asymmetry is not so significant and the blue
and red components of the H$\alpha$ and Ca\,II\,K lines, respectively, 
are not observed separately. Instead, the lines are asymmetric with 
blue or red wing enhancement (Fig.~\ref{ab-fig:allred-2}). 
As in \citet{ab-1999ApJ...521..906A}, the authors 
found that the impulsive flare naturally divides into two phases, an initial 
gentle phase followed by a period of explosive increases in temperature,
pressure and velocity.

\begin{figure}
  \centering
  \mbox{
  \includegraphics[width=5.5cm]{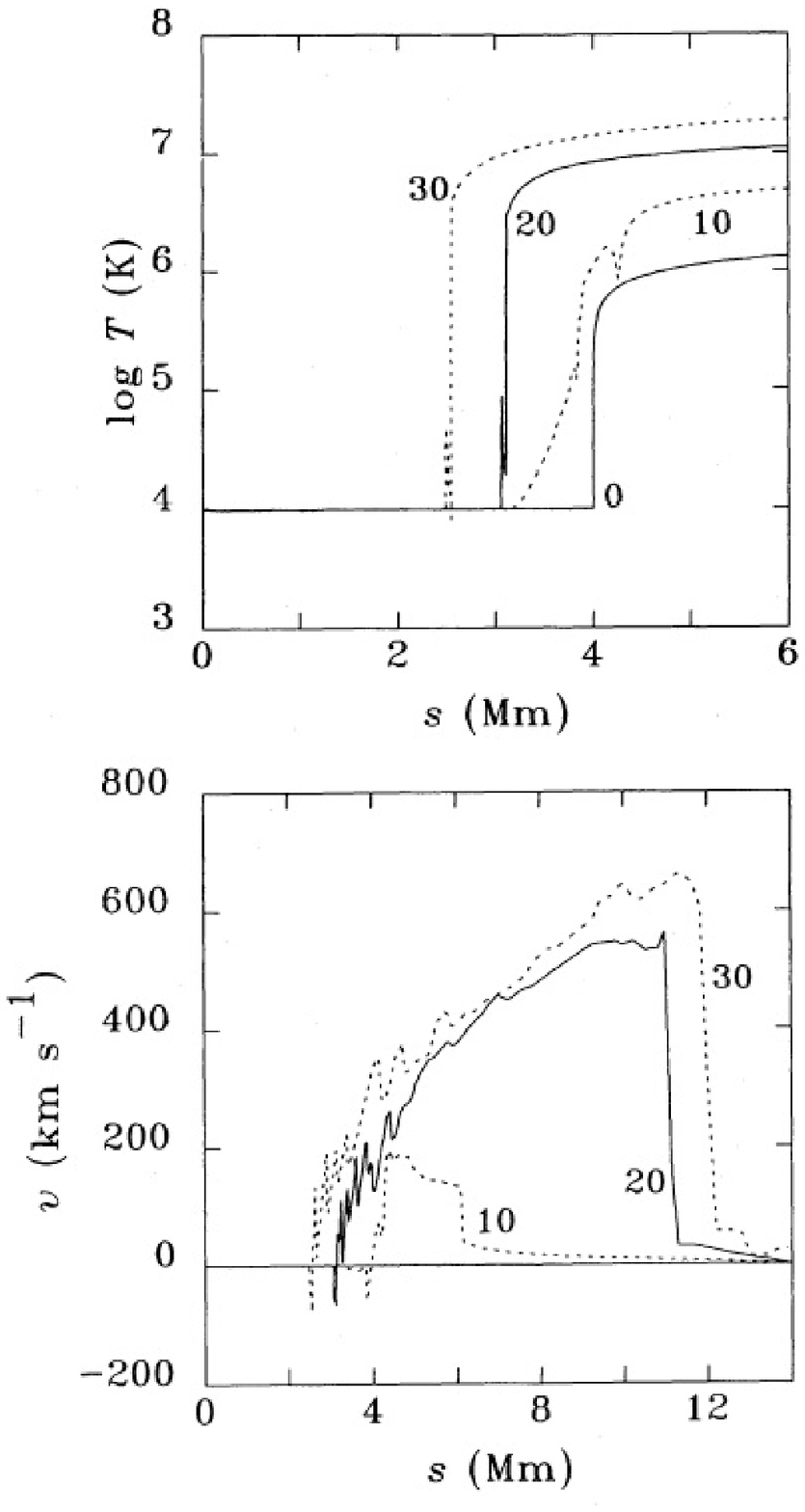}
  \includegraphics[width=7cm]{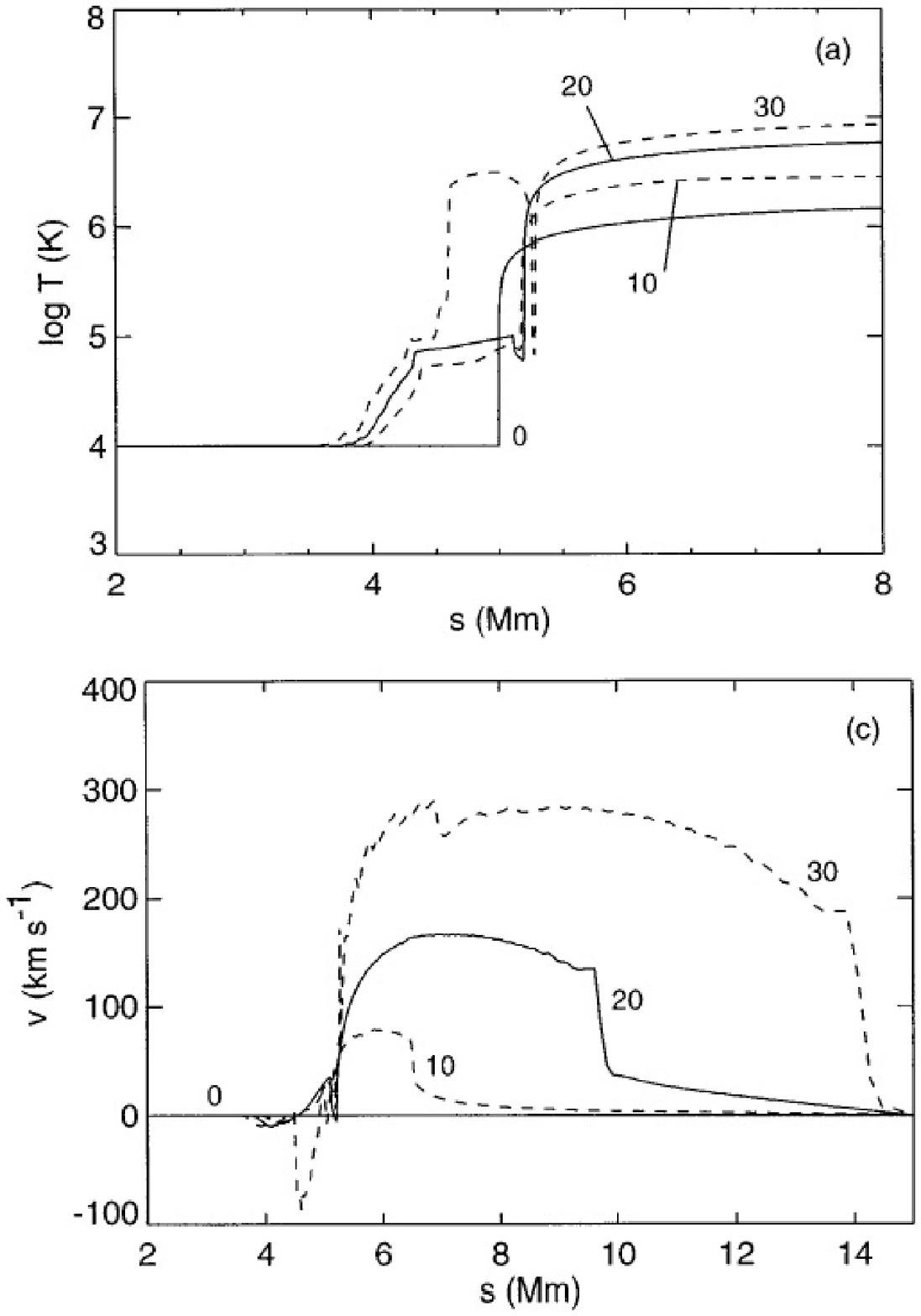}
  }
  \caption[]{Left panels: Hydrodynamic response of the temperature and velocity 
           to the non-thermal electron beam heating (\cite{ab-1989ApJ...341.1067M}) 
           and to the non-thermal proton beam 
           heating (right panels) (\cite{ab-1998ApJ...498..441E}). 
           Positive values of the velocity correspond to upward motion of the plasma. 
           Labels 10, 20, and 30 are the times in seconds after start of non-thermal
           heating.
          }
  \label{ab-fig:mariska-emslie-1}
\end{figure}

There are more papers which treat the problem of plasma flows 
in the non-thermally heated 
chromosphere (\cite{ab-1989ApJ...341.1067M}; \cite{ab-1998ApJ...498..441E}; 
\cite{ab-1992A&A...264..679K}; \cite{ab-1990ApJ...358..328G}; 
\cite{ab-1991A&A...241..618G}). All these hydrodynamic
simulations predict an upflow of the hot coronal plasma due to the enhanced pressure
in the region heated by non-thermal electrons or protons. This upflow is
associated with downflow of chromospheric condensations, but with much
lower velocities (Fig.~\ref{ab-fig:mariska-emslie-1}). These 
condensations disturb the line profiles emitted
from the chromosphere and cause significant asymmetries observed e.g., in
H$\alpha$ line.

The calculated line profiles emerging from flaring atmosphere show roughly
similar behaviour than the observed ones but unfortunately, 
the appearance and the time evolution of the calculated line profiles
was not compared to the spectroscopic observations of particular flares. 
The validity of the modeling summarized in this chapter was not 
confirmed observationally up to now.

\section{Velocity Field in Semiempirical Models of the Flare Atmosphere}

\begin{figure}
  \centering
  \includegraphics[width=8cm]{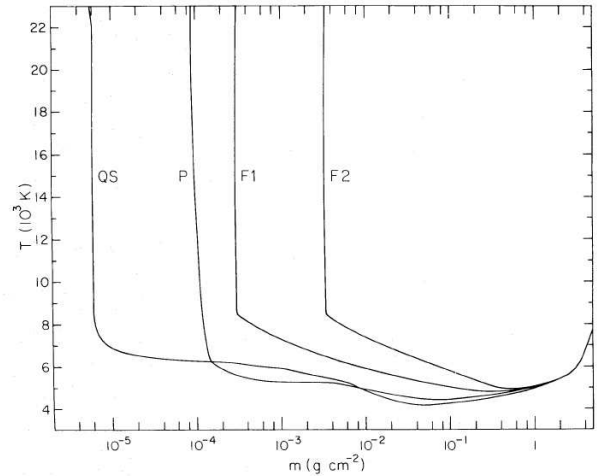}
  \caption[]{Temperature as a function of column mass for the flare models
           F1 and F2, for the quiet-Sun QS (VAL-C) model of \citet{ab-1981ApJS...45..635V} 
           and for the plage (P) model of \citet{ab-1979ApJ...230..924B} 
           (from \cite{ab-1980ApJ...242..336M}).
          }
  \label{ab-fig:mavn}
\end{figure}

Asymmetries observed in chromospheric line profiles are also modeled using
semiempirical flare models. This approach is based on the idea that the temperature
stratification of the atmosphere is determined empirically in the 
way to reproduce the calculated spectrum in the best agreement with observations.
This means that the energy-balance equation is not considered. 
Many spectral lines and continua are used to construct such kind of models. 
Initially, static semiempirical models were developed for a quiet-Sun 
atmosphere (\cite{ab-1981ApJS...45..635V}) and then for the 
static flaring atmosphere (\cite{ab-1980ApJ...242..336M}). 
These models were constructed under assumption of 1-dimensional geometry and hydrostatic
equilibrium (Fig.~\ref{ab-fig:mavn}). They are static however the time sequence of many 
semiempirical models can be used to describe the evolving atmosphere but this method
is valid only for slowly-evolving atmospheres. More detailed description of 
semiempirical models can be found in the paper of Mauas in this book.

Non-LTE radiative transfer methods applied to semiempirical models 
allows us to calculate the spectrum emerging from the atmosphere. In particular,
the profiles of the chromospheric optically thick lines may be 
calculated in details (\cite{ab-1994SoPh..152..393H}). It is also possible to reproduce 
asymmetric line profiles solving the transfer equation for a dynamic 
atmosphere with velocity field, using previously calculated level populations 
for static model (\cite{ab-2005A&A...430..679B}). However, this approach is justified only 
for relatively small velocities ($V \le 10\;\mathrm{km\;s^{-1}}$) which do not 
significantly affect the level populations of the static model (\cite{ab-1998A&AS..127..607N}). 
It cannot be used for impulsive phase of flares 
to model the chromospheric condensations which move quite fast.
Therefore, such simplified calculations are used to model e.g., the gradual phase
of solar flares, when the velocities in the chromosphere are low.

\begin{figure}
  \centering
  \mbox{
  \includegraphics[width=6cm]{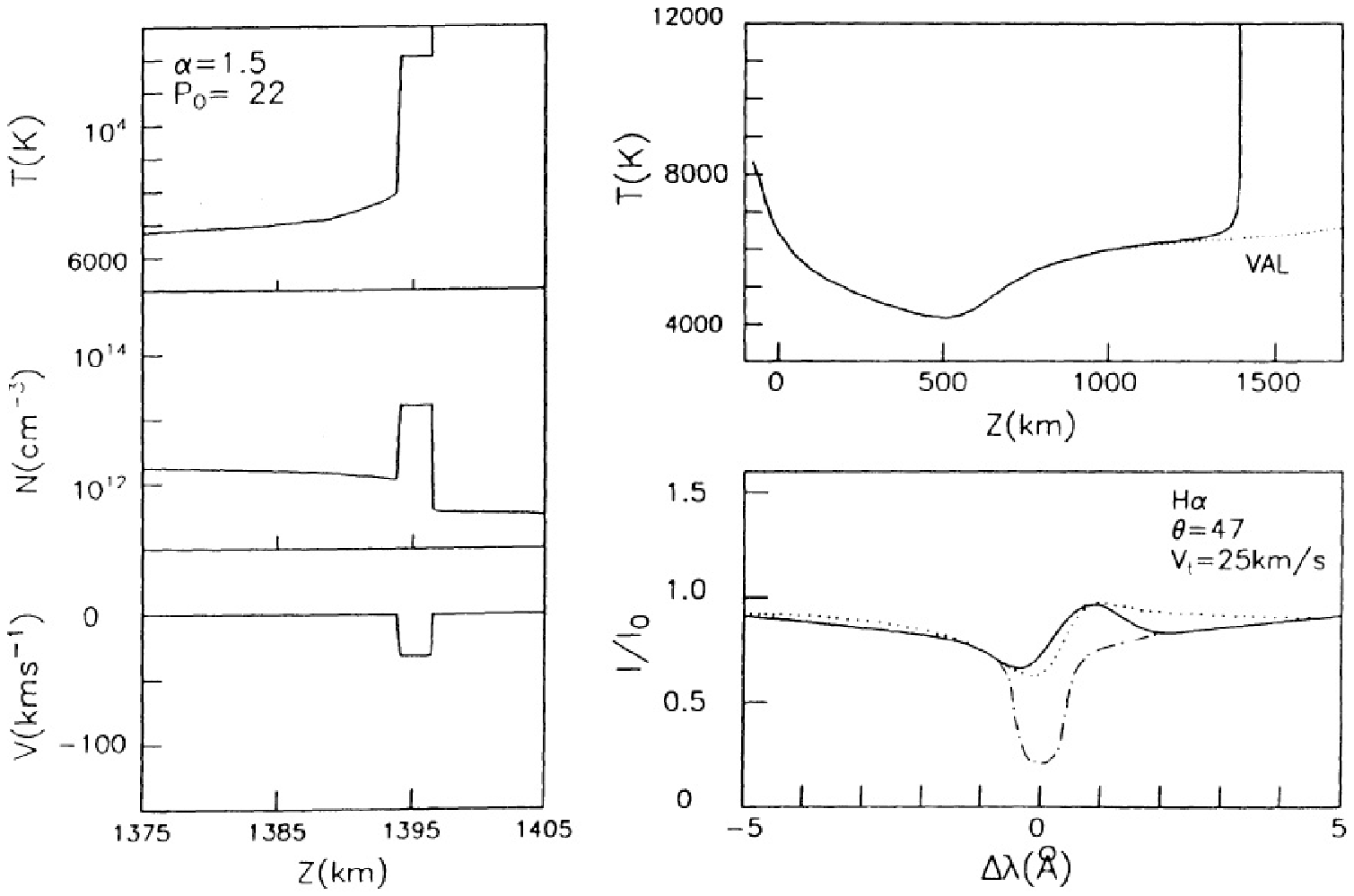}
  \hspace{0.5cm}
  \includegraphics[width=6cm]{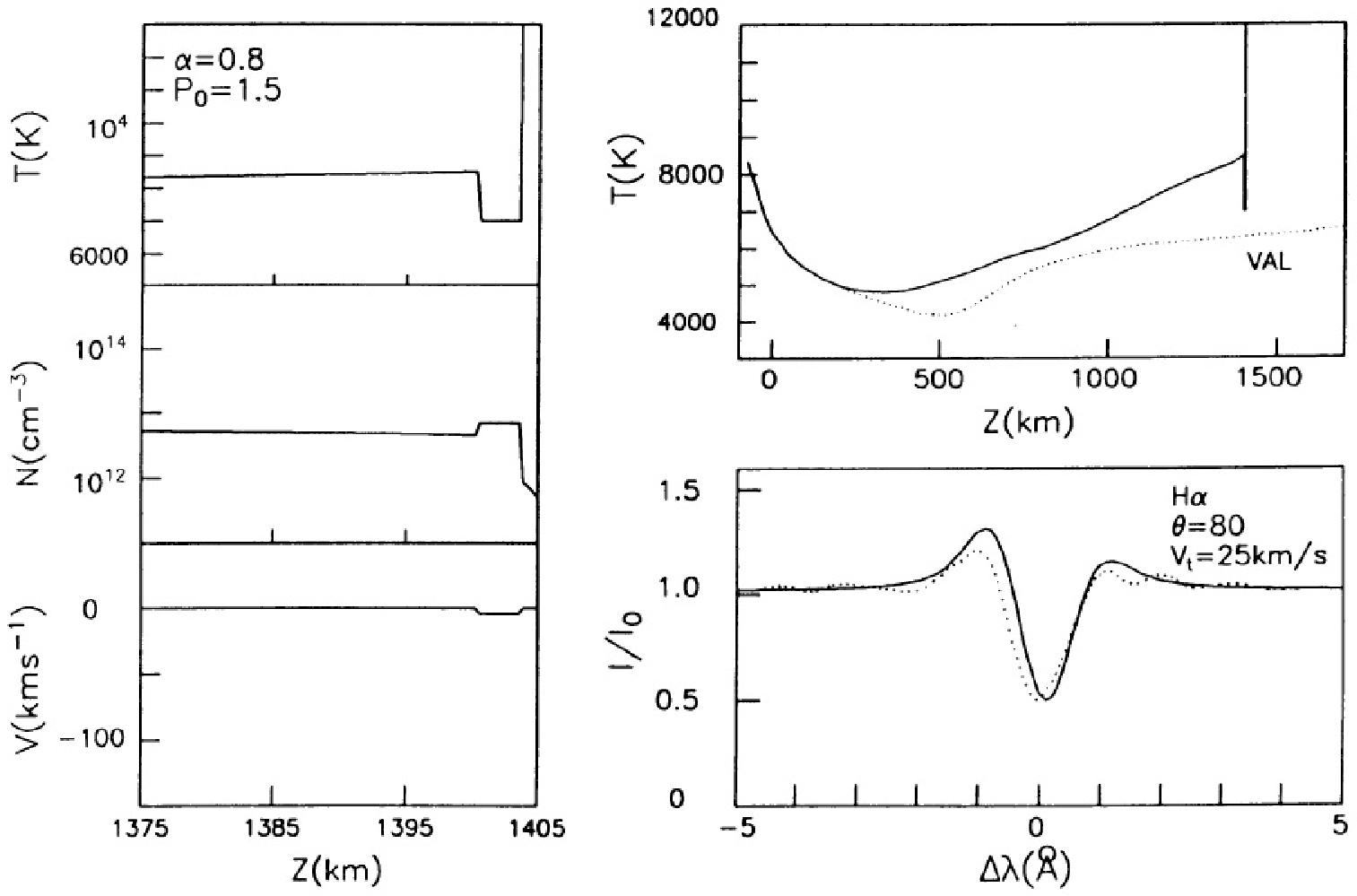}
  }
  \caption[]{Two examples of semiempirical models of the chromosphere with cool condensation.
           Left three panels correspond to the condensation which is responsible
           for the red asymmetry of the H$\alpha$ line while right three panels
           presents condensation producing H$\alpha$ line with blue asymmetry. 
           Temperature, electron density and velocity stratification is presented for
           both cases (negative sign of the velocity corresponds to downflow). 
           The calculated H$\alpha$ line profiles are plotted with 
           solid line, observed ones -- with dotted line (\cite{ab-1993ApJ...416..886G}).
          }
  \label{ab-fig:gan-1}
\end{figure}

One of the first semiempirical modeling of chromospheric 
flows was performed by \citet{ab-1993ApJ...416..886G}. 
Using the H$\alpha$ line profiles observed for two flares the authors
constructed the series of semiempirical models with chromospheric
condensations. It was shown that chromospheric condensations are responsible
not only for the red asymmetry of the H$\alpha$ line, but also for the
blue asymmetry of the line with central reversal (Fig.~\ref{ab-fig:gan-1}). 
Chromospheric condensations were assumed to be homogeneous with constant
velocity. The most important conclusion of this paper is that the properties
of chromospheric condensations seem to be consistent with the results of 
hydrodynamical models of solar flares. Comparison of calculated H$\alpha$
line profiles with real observations present also a valuable part of this paper.

\begin{figure}
  \centering
  \includegraphics[width=5cm]{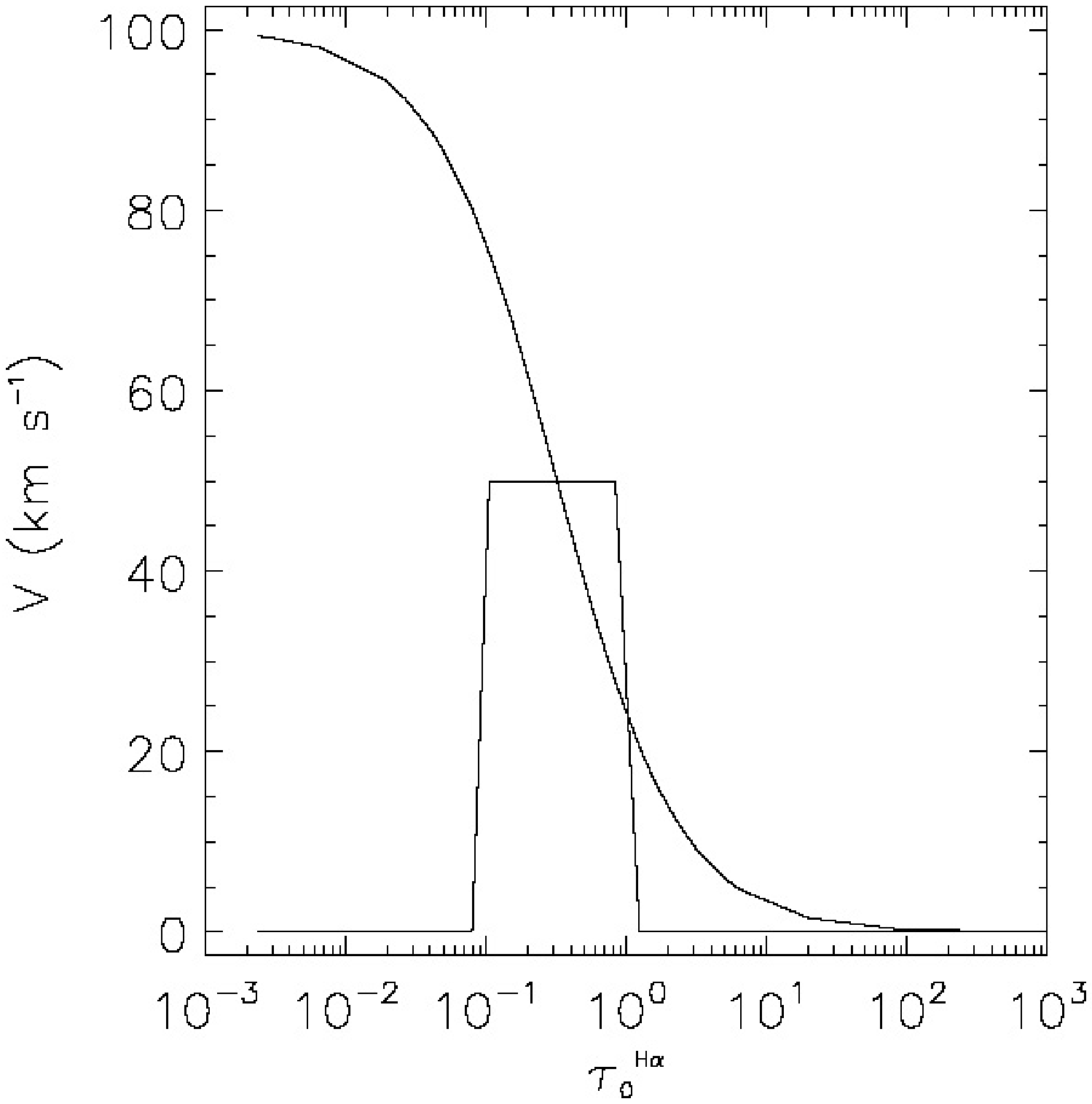}
  \includegraphics[width=6cm]{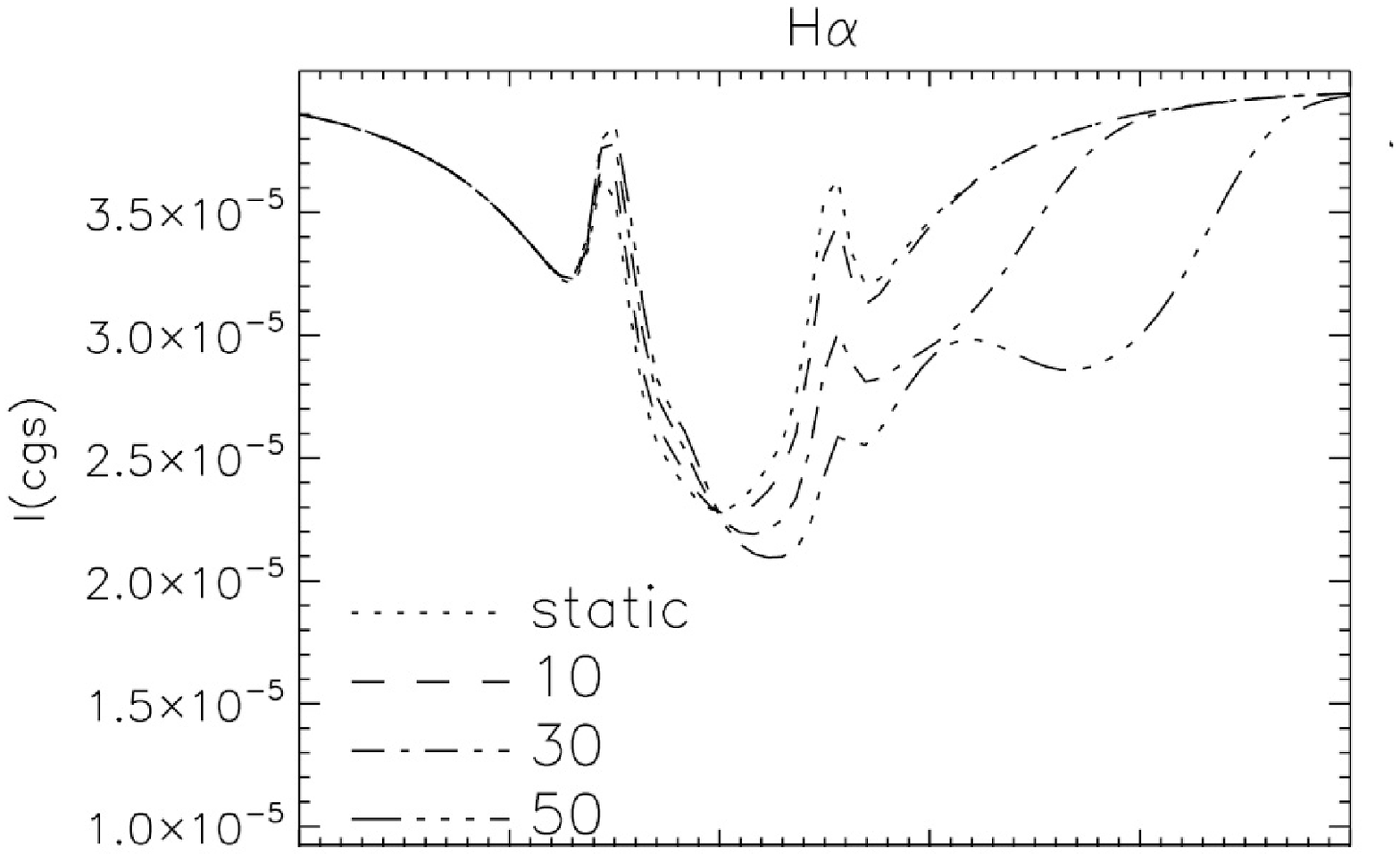}
  \vspace{1cm}
  \caption[]{{\em Left:} Models of the velocity field in the chromosphere used in
           the non-LTE calculations (positive sign of the velocity corresponds 
           to downflow). {\em Right:} Some examples of synthetic H$\alpha$ 
           line profiles calculated for a static and dynamic atmosphere with different
           velocities 10, 30, and 50\,km\,s$^{-1}$ (\cite{ab-1998A&AS..127..607N}).
          }
  \label{ab-fig:nejezchleba-1}
\end{figure}

An interesting work was presented by \citet{ab-1998A&AS..127..607N} who simulate the
influence of the velocity field on the H$\alpha$ line profiles. The calculations
were performed using non-LTE model of plane-parallel solar flare atmosphere
with stationary velocity field. This velocity field was applied to different
layers of the solar atmosphere and the emergent H$\alpha$ line profiles
were calculated for two models of solar flare F1 and F2 (\cite{ab-1980ApJ...242..336M}).
Figure~\ref{ab-fig:nejezchleba-1} present an example of asymmetric line profiles
calculated for a weak-flare model atmosphere F1 and for different velocity
fields. The main conclusion of this work is that the velocity
field affects the level populations via the increase of the downward radiation.
Nevertheless, for velocities that do not exceed the thermal velocity of plasma, 
one can use the static populations for the formal solution of radiative 
transfer equation including the velocity to reproduce the observed line 
asymmetries. Other important point is that application of the bisector 
method would lead in some cases to reverse velocity, in others to underestimation
of the velocity. It includes, besides the part of the profile directly affected 
by the moving material, also a ``static'' part of the profile. To use the 
bisector in terms of Doppler-shift the static part should be somehow
eliminated. This remark makes questionable all estimations of
the Doppler velocity obtained with the bisector method applied to self-reversed
or emission chromospheric lines observed in solar flares.

The conclusions of \citet{ab-1998A&AS..127..607N} suggest that deducing the velocity from 
flare line profiles is rather difficult and cannot be done only by searching for 
Doppler-shifts with the bisector method. Chromospheric line profiles suggest that 
the flare atmosphere is highly dynamic and stratified with rather complicated plasma 
motion. These lines are optically ``thick'' and the 
only reliable way to analyze the flows is to use the non-LTE 
radiative transfer codes, which enable us to compute the chromospheric models with 
velocity fields. Resulting synthetic line profiles can then be compared with the 
observed ones.

\begin{figure}
  \centering
  \mbox{
  \includegraphics[width=6cm]{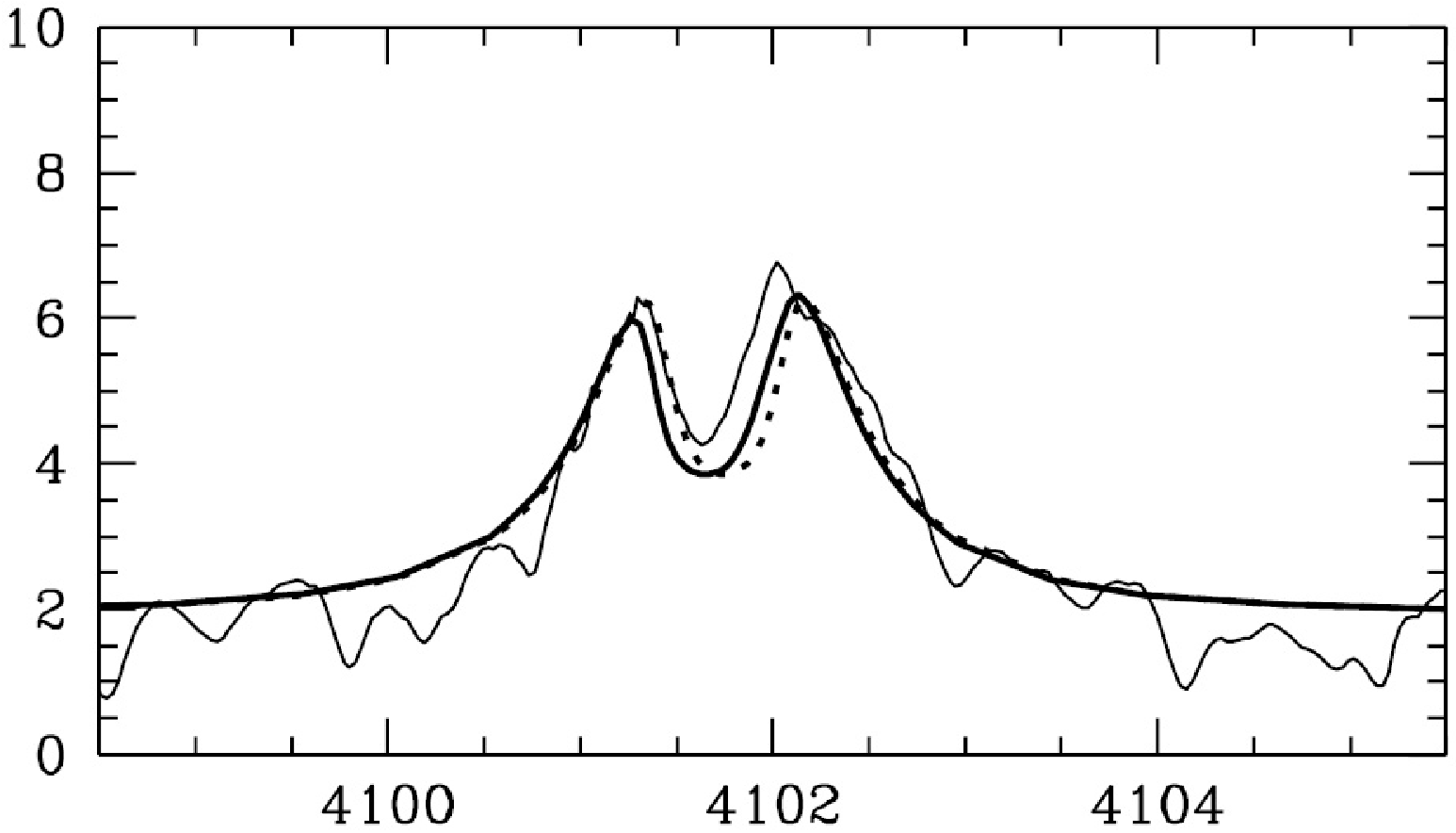}
  \includegraphics[width=6.5cm]{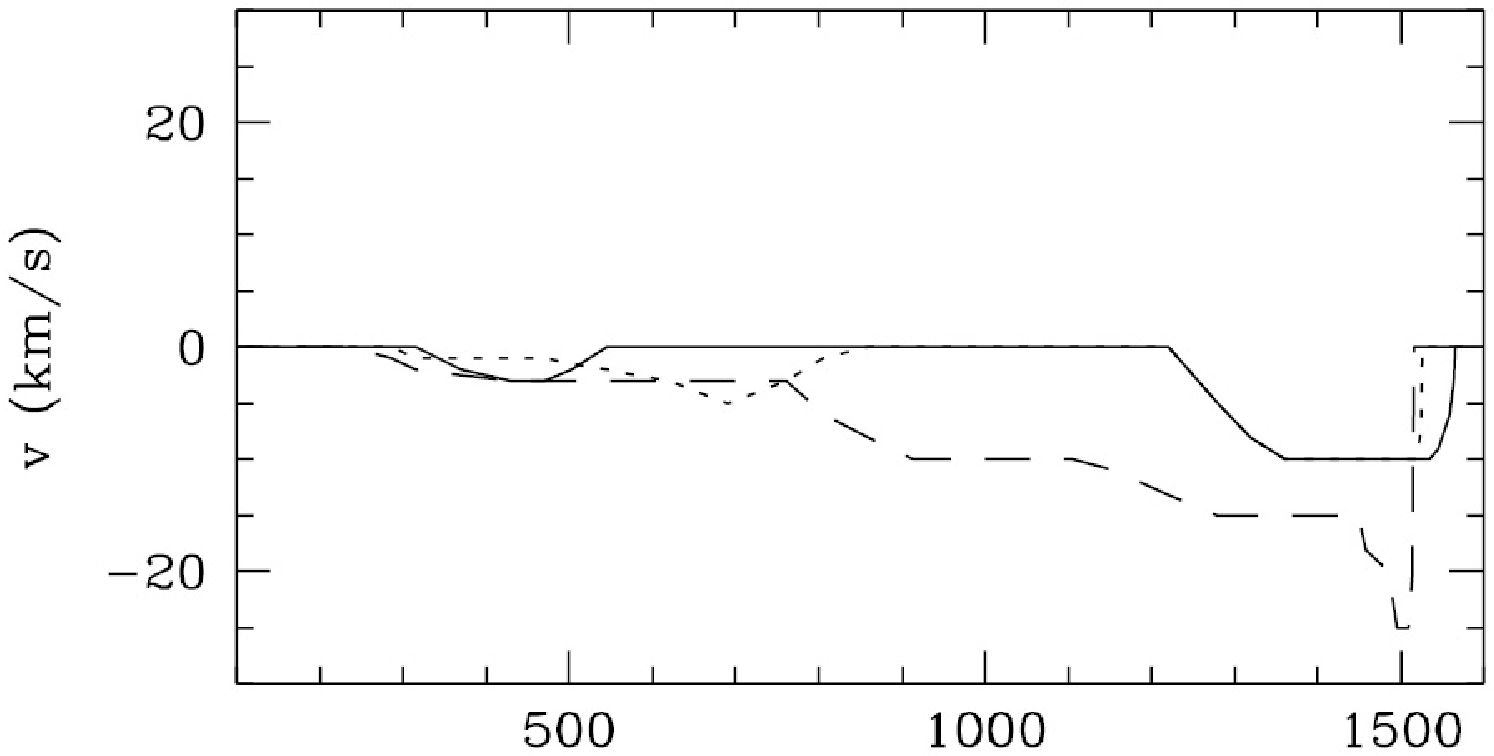}
  }
  \includegraphics[width=8cm]{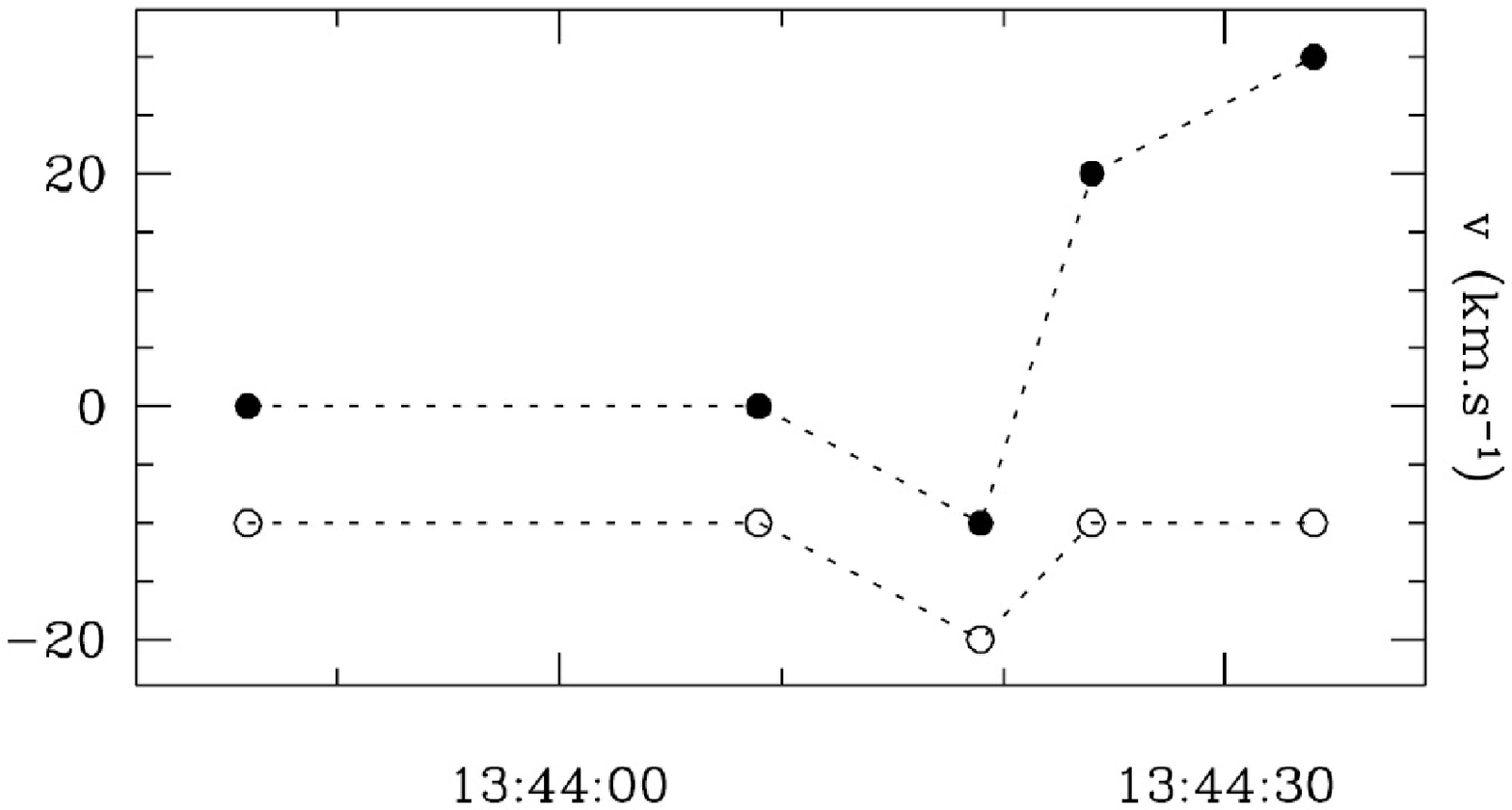}
  \caption[]{An example of the observed (thin line) and fitted (thick line) H$\delta$ line 
           profiles and determined velocity field across the atmosphere. Negative 
           value indicates an upward velocity. In the lower
           panel the time evolution of the velocity value at two different heights in 
           the atmosphere is presented. Filled circles refer to the height $\approx$~900 
           and open circles to the height $\approx 1400$\,km (\cite{ab-2002A&A...387..678F}).
          }
  \label{ab-fig:falchi-mauas-1}
\end{figure}

The direct comparison of the observed and synthetic line profiles was
presented in the paper of \citet{ab-2002A&A...387..678F}. They study the chromospheric 
structure of a small flare and construct 5 semiempirical models for different times, 
which reproduce the profiles of the H$\delta$, Ca\,II\,K, and
Si\,I~3905\,\AA\ lines during the flare evolution. In order to reproduce the asymmetry
of the lines the velocity fields were introduced in the line profile calculations. 
The modeling was done using the non-LTE Pandora code of \citet{ab-1984mrt..book..341A}.
The trial-and-error method was used to reproduce the observed line profiles
by the synthetic ones. Figure~\ref{ab-fig:falchi-mauas-1} (upper panels) present an example
of the observed and fitted H$\delta$ line profiles and determined velocity 
field across the atmosphere. In the lower panel of Fig.~\ref{ab-fig:falchi-mauas-1}
the time evolution of the velocity value at two different heights in the atmosphere
is presented. The presence of an upward motion in the flaring atmosphere at 1400\,km, 
might be a signature of the chromospheric evaporation observed at chromospheric levels.
It is interesting to notice that around 13:44:30 UT the downflow is observed at
the height of 900\,km, while the upward motion is evident at 1400\,km above the 
photosphere. One possible explanation is that chromospheric evaporation together
with condensations is observed. In this case chromospheric evaporation is observed 
at chromospheric levels and not, as more common, at coronal levels.

\begin{figure}
  \centering
  \includegraphics[width=12cm]{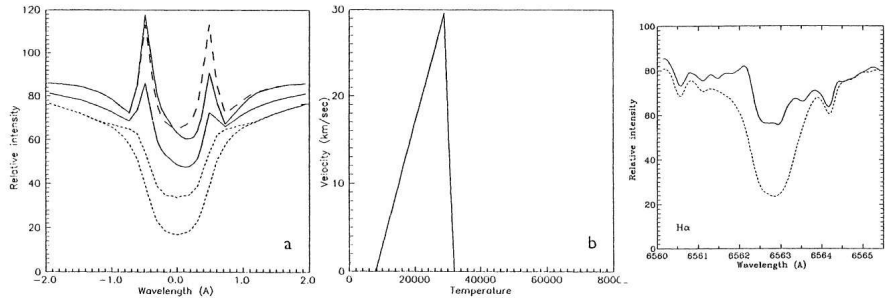}
  \caption[]{Left: Simulation of the influence of the downward velocity field
           on the emergent H$\alpha$ line profile emitted from flaring
           model atmosphere F1 of Machado et al.\ (1980). In the middle panel
           the velocity used in the modeling as a function of temperature is shown 
           (positive sign of the velocity indicates downflow).
           Also an example of the observed H$\alpha$ line with red asymmetry is 
           presented (for details see \citet{ab-1994SoPh..152..393H}).
  }
  \label{ab-fig:heinzel-1}
\end{figure}

Using the semiempirical F1 model of a weak 
flare (\cite{ab-1980ApJ...242..336M}), \citet{ab-1994SoPh..152..393H} 
showed that the blue asymmetry of H$\alpha$ line profile is observed
due to the downflow of chromospheric plasma. It is interesting to notice that
the blue asymmetry is associated with the red-shift of the central
absorption feature. Similar results was shown by \citet{ab-1993ApJ...416..886G}.
The structure of the velocity field
used in the non-LTE simulations of \citet{ab-1994SoPh..152..393H} 
was qualitatively consistent with the concept
of downward-moving chromospheric condensations (Fig.~\ref{ab-fig:heinzel-1}).
These calculations were performed using the non-LTE code developed by 
\citet{ab-1995A&A...299..563H} and modified for flare conditions. 
The code uses a 1D plane-parallel geometry and the
atmosphere is in hydrostatic equilibrium. Hydrogen excitation and 
ionization equilibrium have been computed by solving simultaneously the
radiative transfer equations, the equations of statistical equilibrium
for a five-level plus continuum atomic model of hydrogen and the
equations of particle and charge conservation. The equations of
statistical equilibrium have been preconditioned according to
\citet{ab-1991A&A...245..171R}. The preconditioning is based on the
lambda-operator splitting technique, where the exact lambda operator
is expressed as an approximate operator plus the correction.
Then the correction is iteratively applied to a lagged source function by using 
the so-called Accelerated Lambda Iterations (ALI) method. 
For multilevel atoms this method is referred to as MALI -- Multilevel Accelerated
Lambda Iterations (\cite{ab-1991A&A...245..171R}). The preconditioned 
equations are then linearized with respect to the atomic level populations 
and electron density and solved iteratively (\cite{ab-1995A&A...299..563H}). 

\begin{figure}
  \centering
  \includegraphics[width=9.5cm]{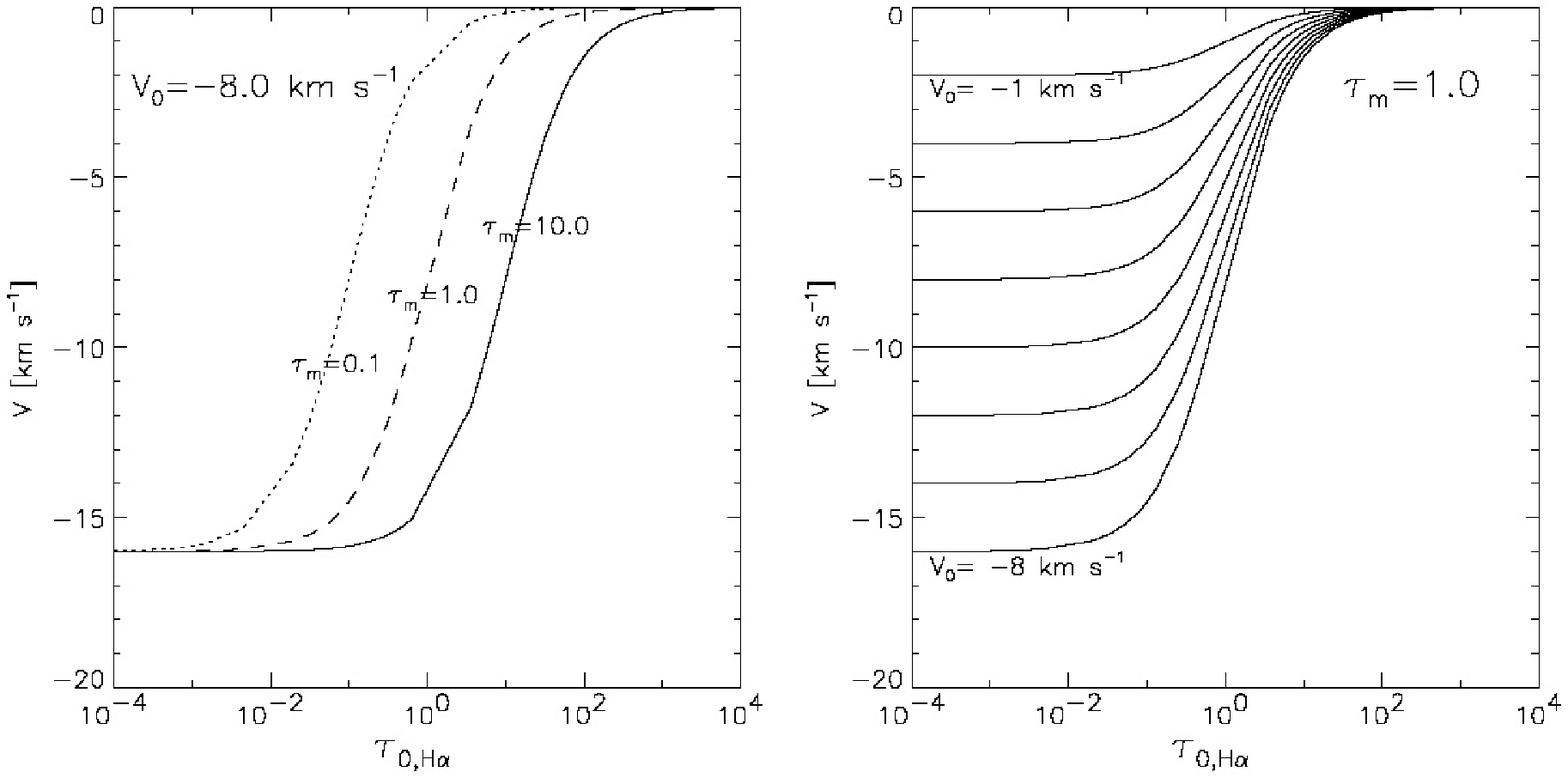}
  \includegraphics[width=10cm]{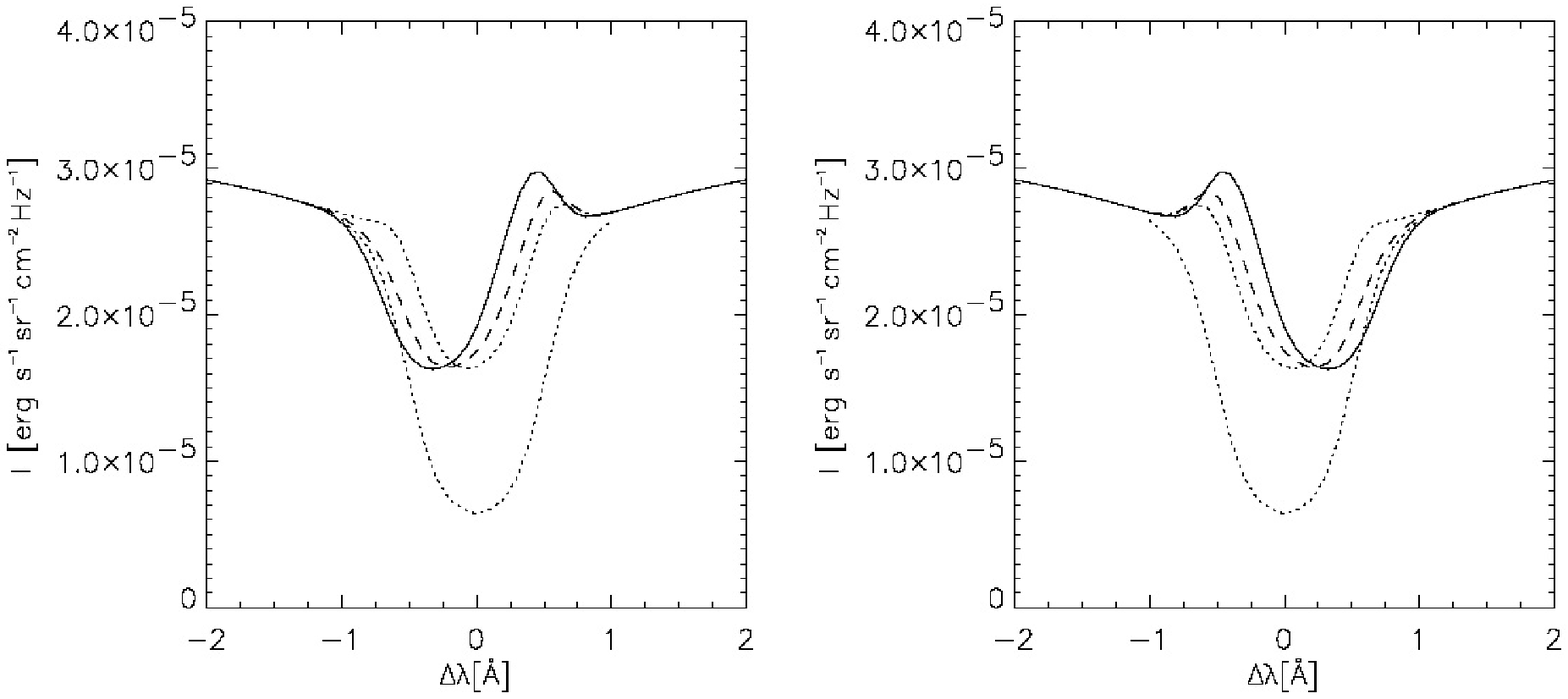}
  \caption[]{The H$\alpha$ line-centre optical depth distribution of 
           the velocity field used in the modeling plotted for different values 
           of $\tau_{\mathrm{m}}$ and for $V_{0}=-8\;\mathrm{km\;s^{-1}}$
          (upper left panel) and for different values of $V_{\mathrm{0}}$ and for
          $\tau_\mathrm{m}=1.0$ (upper right panel). In the lower panels
          the influence of the velocity field on H$\alpha$ line profiles
          emitted from flare is presented (upflow defined by $V_{0}=-8$ 
          and downflow $V_{0}=+8\;\mathrm{km\;s^{-1}}$ for three values of 
	        $\log \tau_{\rm m} = 0.1, 1.0$, and 10.0
          (dotted, dashed and continuous lines, respectively) (\cite{ab-2005A&A...430..679B}).
  }
  \label{ab-fig:berlicki-1}
\end{figure}

This non-LTE code was also used in \citet{ab-2005A&A...430..679B} to analyse
the time evolution of the line asymmetry observed during the gradual
phase of the solar flare on October 22, 2002. In this paper for the first 
time the evaporative flows in the gradual phase are studied 
quantitatively by using a non-LTE radiative transfer code and
spectroscopic observations of the flare ribbons. First the authors analyse
the influence of different velocity fields on the emergent H$\alpha$
line profile. Again, it was shown that the downflow of flaring plasma
causes blue asymmetry of the self-reversed line while upflow -- 
red-asymmetry (Fig.~\ref{ab-fig:berlicki-1}). For the modeling of the 
observed line asymmetry except the changes of the value of the velocity,
also the height of the velocity field in the atmosphere was different.
The procedure of fitting the H$\alpha$ line profiles was performed 
using a grid of many models by varying different parameters.
Each observed profile was fitted by the least-square technique 
to a closest synthetic profile from the grid and the model with
the velocity field was found for each analysed line profile.

\begin{figure}
  \centering
  \mbox{
  \includegraphics[width=6cm]{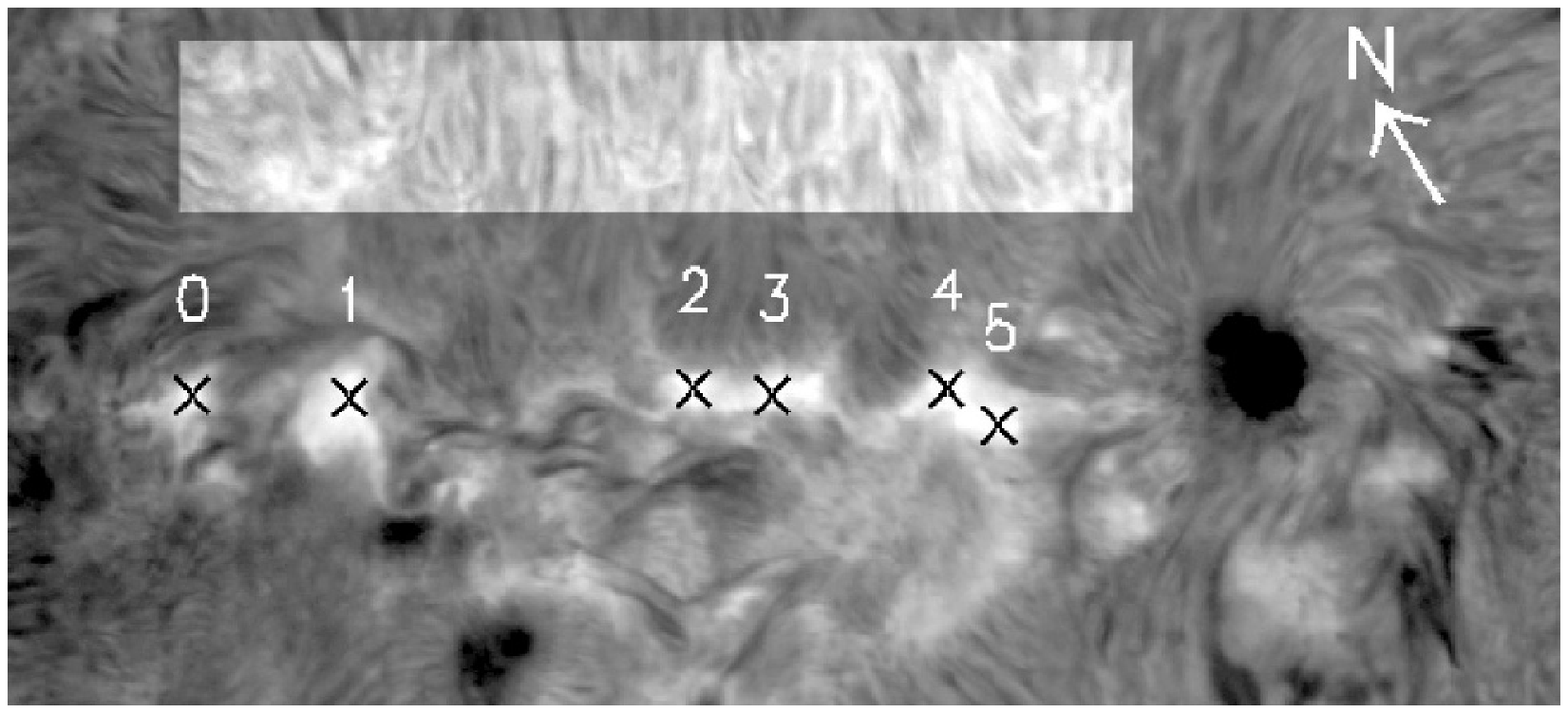}
  \includegraphics[width=6cm]{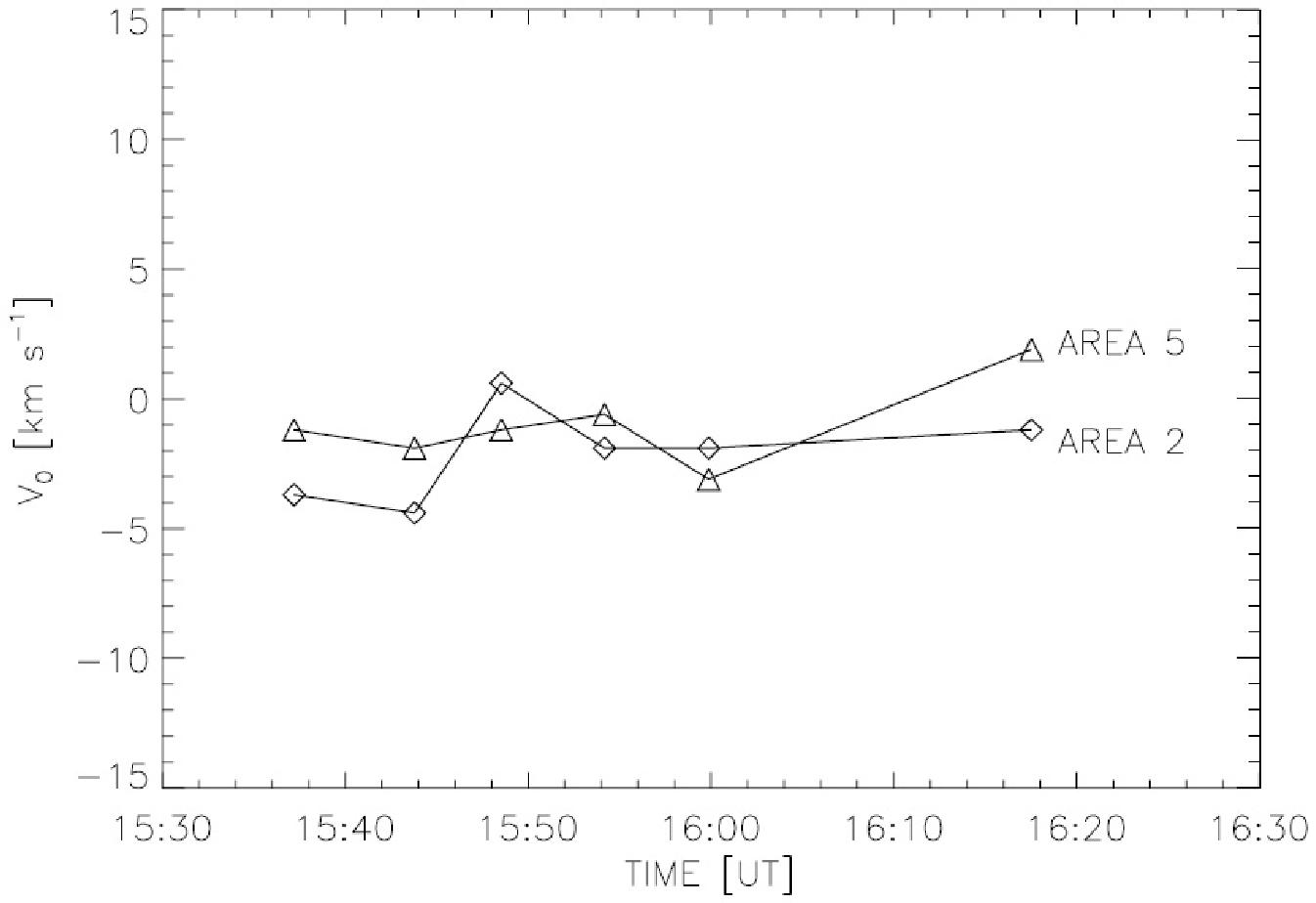}
  }
  \includegraphics[width=11cm]{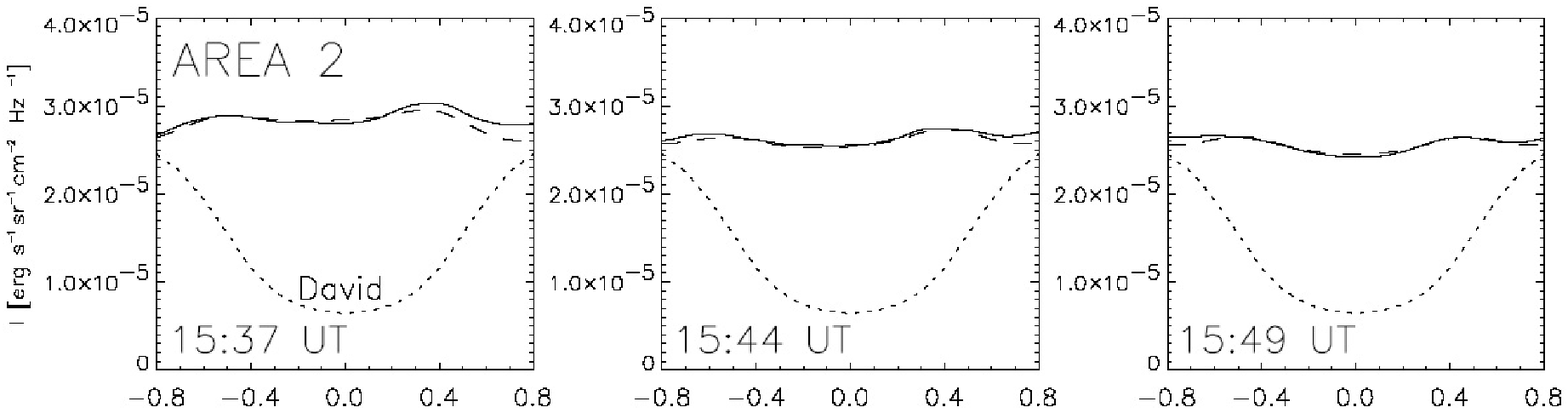}
  \caption[]{The image of the flare on October 22, 2002 used in the
           analysis of \citet{ab-2005A&A...430..679B} and the observed (continuous
           lines) and fitted (dashed lines) H$\alpha$ line profiles (upper and lower
           left panels). 0~--~5 are the areas used in the analysis. The 
           time evolution of the velocity in the chromosphere deduced from 
           line asymmetries is plotted for two analysed areas in the right panel. 
           Negative velocities correspond to upflow (\cite{ab-2005A&A...430..679B}).
  }
  \label{ab-fig:berlicki-2}
\end{figure}

In the analysis the MSDP (Multichannel Subtractive Double Pass) 
spectrograph (Mein 1991) coupled to the VTT telescope working
at the Teide Observatory (Tenerife, Canary Islands) was used.
36 H$\alpha$ line profiles (six areas at six different
times) observed during the M1.0 flare on October 22, 2002 were taken 
for the analysis (Fig.~\ref{ab-fig:berlicki-2} -- upper left panel). 
As an example we present in Fig.~\ref{ab-fig:berlicki-2} 
(lower panel) some profiles observed in chosen area at three times.
These observed profiles (solid lines) are fitted with the 
synthetic ones (dashed lines)
obtained from the grid. In the right panel of Fig.~\ref{ab-fig:berlicki-2}
the temporal evolution of the velocity field is presented for
two different areas of the flare.

The authors interpret the upflows found in the flare ribbons in terms
of the \citet{ab-1978ApJ...220.1137A} model for gentle evaporation.
This process may occur during the gradual phase of
solar flares and it can be driven by conductive heat flux from
the high-temperature flare plasma contained in magnetic flux
tubes above the photosphere. In the future it would
be interesting to use more spatial points at more times and to
use the spectra obtained within a wider range of wavelengths.
Other distributions of the velocity field in the chromosphere
should also be tested. In addition, to perform non-LTE modeling
of the flare structure it would be useful to have other
spectral lines formed at different levels of the chromosphere.

\section{Summary}

In this review I presented some interesting papers concerning plasma
flows observed during solar flares in cool chromospheric layers.
These flows are directly responsible for the line-asymmetries 
and/or line-shifts often observed in chromospheric lines emitted
by the flaring plasma. 

An important work was done to understand the flows and their mechanisms.
In order to determine the plasma velocity and flow direction the
bisector method was applied for line profiles. Unfortunately, as
we could see, this method leads in some cases to misleading
estimations of the velocity. Recently, the direct comparison
of the observed and synthetic line profiles gives more
valuable information about velocity fields in the chromosphere.
All the data support the evaporative model of solar flares where
explosive chromospheric evaporation of the hot plasma is
associated with the chromospheric condensations observed
in ``cool'' chromospheric lines. In the late phases of flares
the gentle evaporation may be observed in chromospheric
lines. 

For the future it is necessary to use large and dense grids of 
the chromospheric models computed with hydrodynamic and non-LTE codes. 
They may help us to understand the flows and give more realistic 
description of the physical processes
during the flares, particularly the heating mechanisms and their
role at different phases of the flare evolution.

Finally, really good spectral observations of flares are needed.
They have to be co-spatial, simultaneous and obtained in different 
spectral ranges (X-ray, EUV, UV, optical, IR). Such observations 
would be very helpful to construct the full picture of the plasma 
flows during flares. There are some data concerning the flows observed
in soft X-ray and EUV but they are extremely rare and almost never 
cospatial nor simultaneous with the observations in chromospheric lines.\\

\acknowledgements 
  This research was supported by the European Commission 
  through the RTN programme (European Solar Magnetism Network, 
  contract HPRN-CT-2002-00313. The author also would like to
  thank P. Heinzel for helpful comments and valuable remarks.


\end{document}